\let\savedbibliographystyle\bibliographystyle
\renewcommand{\bibliographystyle}[1]{}
\PassOptionsToPackage{hypertexnames=false}{hyperref}
\documentclass{optica-article}
\ifdefined\OEtitle
  \let\bibliographystyle\savedbibliographystyle
\fi
\journal{opticajournal}
\articletype{Research Article}
\usepackage{amsmath,array,booktabs,caption,placeins,mathrsfs}
\usepackage{lineno}
\usepackage{chapterbib}
\graphicspath{{figures/}}
\hypersetup{colorlinks=true,linkcolor=urlblue,filecolor=urlblue,
  citecolor=urlblue,urlcolor=urlblue}

\begin{document}
\ifdefined\OEtitle\else
  \def\htmlbibliography#1{\begin{thebibliography}{10}
\newcommand{\enquote}[1]{``#1''}

\bibitem{VanCittert1934}
P.~H. van Cittert, \enquote{{Die wahrscheinliche Schwingungsverteilung in einer
  von einer Lichtquelle direkt oder mittels einer Linse beleuchteten Ebene},}
  \href{https://doi.org/10.1016/S0031-8914(34)90026-4}{{\protect\JournalTitle{Physica}}}
  \textbf{1}, 201--210 (1934).

\bibitem{Zernike1938}
F.~Zernike, \enquote{{The concept of degree of coherence and its application to
  optical problems},}
  \href{https://doi.org/10.1016/S0031-8914(38)80203-2}{{\protect\JournalTitle{Physica}}}
  \textbf{5}, 785--795 (1938).

\bibitem{Goodman1985}
J.~W. Goodman,
  \href{https://books.google.com/books/about/Statistical_Optics.html?id=2VTwAAAAMAAJ}{\emph{{Statistical
  Optics}}} (Wiley, 1985).

\bibitem{MandelWolf1965}
L.~Mandel and E.~Wolf, \enquote{{Coherence Properties of Optical Fields},}
  \href{https://doi.org/10.1103/RevModPhys.37.231}{{\protect\JournalTitle{Reviews
  of Modern Physics}}} \textbf{37}, 231--287 (1965).

\bibitem{Hopkins1951}
H.~H. Hopkins, \enquote{{The concept of partial coherence in optics},}
  \href{https://doi.org/10.1098/rspa.1951.0158}{{\protect\JournalTitle{Proceedings
  of the Royal Society of London. Series A}}} \textbf{208}, 263--277 (1951).

\bibitem{Mehta2014}
S.~B. Mehta and R.~Oldenbourg, \enquote{{Image simulation for biological
  microscopy: microlith},}
  \href{https://doi.org/10.1364/BOE.5.001822}{{\protect\JournalTitle{Biomedical
  Optics Express}}} \textbf{5}, 1822--1838 (2014).

\bibitem{Schell1967}
A.~C. Schell, \enquote{{A technique for the determination of the radiation
  pattern of a partially coherent aperture},}
  \href{https://doi.org/10.1109/TAP.1967.1138864}{{\protect\JournalTitle{IEEE
  Transactions on Antennas and Propagation}}} \textbf{15}, 187--188 (1967).

\bibitem{Nugent1990}
K.~A. Nugent, \enquote{{A generalization of Schell's theorem},}
  \href{https://doi.org/10.1016/0030-4018(90)90066-3}{{\protect\JournalTitle{Optics
  Communications}}} \textbf{79}, 267--269 (1990).

\bibitem{Montoya2025}
M.~Montoya, M.~J. Lopera, Y.~Nie, \emph{et~al.}, \enquote{{Generalized Van
  Cittert-Zernike Schell propagator: an efficient algorithm for simulating
  partially coherent light},}
  \href{https://doi.org/10.1364/OE.571673}{{\protect\JournalTitle{Optics
  Express}}} \textbf{33}, 46313--46325 (2025).

\bibitem{Harvey1999}
J.~E. Harvey, C.~L. Vernold, A.~Krywonos, \emph{et~al.}, \enquote{{Diffracted
  radiance: a fundamental quantity in nonparaxial scalar diffraction theory},}
  \href{https://doi.org/10.1364/AO.38.006469}{{\protect\JournalTitle{Applied
  Optics}}} \textbf{38}, 6469--6481 (1999).

\bibitem{Harvey2000Errata}
J.~E. Harvey, C.~L. Vernold, A.~Krywonos, \emph{et~al.}, \enquote{{Diffracted
  radiance: a fundamental quantity in nonparaxial scalar diffraction theory:
  errata},}
  \href{https://doi.org/10.1364/AO.39.006374}{{\protect\JournalTitle{Applied
  Optics}}} \textbf{39}, 6374--6375 (2000).

\bibitem{HarveyVernold1998}
J.~E. Harvey and C.~L. Vernold, \enquote{{Description of Diffraction Grating
  Behavior in Direction Cosine Space},}
  \href{https://doi.org/10.1364/AO.37.008158}{{\protect\JournalTitle{Applied
  Optics}}} \textbf{37}, 8158--8159 (1998).

\bibitem{Muellers2022}
D.~M{\"u}llers, J.~Kuhl, and S.~Kontermann, \enquote{{Nonparaxial optical
  transfer function for arbitrary illumination in partially coherent imaging
  systems and the oblique source application},}
  \href{https://doi.org/10.1364/JOSAA.452462}{{\protect\JournalTitle{Journal of
  the Optical Society of America A}}} \textbf{39}, 744--758 (2022).

\bibitem{Peng2021}
Y.~Peng, S.~Choi, J.~Kim, \emph{et~al.}, \enquote{Speckle-free holography with
  partially coherent light sources and camera-in-the-loop calibration,}
  \href{https://doi.org/10.1126/sciadv.abg5040}{{\protect\JournalTitle{Science
  Advances}}} \textbf{7}, eabg5040 (2021).

\bibitem{Montoya2026}
M.~Montoya, Y.~Nie, and D.~Blinder, \enquote{{Computer-generated holography
  using the generalized Van Cittert-Zernike Schell propagator},}
  \href{https://doi.org/10.1364/OL.595741}{{\protect\JournalTitle{Optics
  Letters}}} \textbf{51}, 2084--2087 (2026).

\bibitem{Sinha2026}
R.~K. Sinha and M.~Mojahedi, \enquote{Nonlocal leaky mode metasurfaces for
  mode-, angle-, and polarization-multiplexed holography with thermal light,}
  \href{https://doi.org/10.1002/lpor.72007}{{\protect\JournalTitle{Laser \&
  Photonics Reviews}}} p. e72007 (2026).

\bibitem{Poulsen2025}
K.~E.~S. Poulsen, X.~Zambrana-Puyalto, R.~de~la Fuente~Herrezuelo,
  \emph{et~al.}, \enquote{Experimental demonstration of a beam shaping
  non-imaging metasurface,}
  \href{https://doi.org/10.1364/OE.559542}{{\protect\JournalTitle{Optics
  Express}}} \textbf{33}, 19119--19129 (2025).

\bibitem{Khaidarov2020}
E.~Khaidarov, Z.~Liu, R.~Paniagua-Dom{\'i}nguez, \emph{et~al.},
  \enquote{Control of {LED} emission with functional dielectric metasurfaces,}
  \href{https://doi.org/10.1002/lpor.201900235}{{\protect\JournalTitle{Laser \&
  Photonics Reviews}}} \textbf{14}, 1900235 (2020).

\bibitem{Ellingen2025}
E.~Ellingen and R.~Lange, \enquote{{Beam Shaping of VCSEL Arrays Using
  Randomized Microlens Arrays With Integrated Collimation},}
  \href{https://doi.org/10.1109/JPHOT.2025.3603888}{{\protect\JournalTitle{IEEE
  Photonics Journal}}} \textbf{17}, 2201209 (2025).

\bibitem{Kim2020}
D.~C. Kim, A.~Hermerschmidt, P.~Dyachenko, \emph{et~al.}, \enquote{{Inverse
  design and demonstration of high-performance wide-angle diffractive optical
  elements},}
  \href{https://doi.org/10.1364/OE.394442}{{\protect\JournalTitle{Optics
  Express}}} \textbf{28}, 22321--22333 (2020).

\bibitem{Wu2024}
C.~Wu, X.~Huang, Y.~Ji, \emph{et~al.}, \enquote{{Addressable structured light
  system using metasurface optics and an individually addressable VCSEL
  array},}
  \href{https://doi.org/10.1364/PRJ.516942}{{\protect\JournalTitle{Photonics
  Research}}} \textbf{12}, 1129--1137 (2024).

\bibitem{Bluestein1970}
L.~Bluestein, \enquote{{A linear filtering approach to the computation of
  discrete Fourier transform},}
  \href{https://doi.org/10.1109/TAU.1970.1162132}{{\protect\JournalTitle{IEEE
  Transactions on Audio and Electroacoustics}}} \textbf{18}, 451--455 (1970).

\bibitem{MatsushimaShimobaba2009}
K.~Matsushima and T.~Shimobaba, \enquote{{Band-limited angular spectrum method
  for numerical simulation of free-space propagation in far and near fields},}
  \href{https://doi.org/10.1364/OE.17.019662}{{\protect\JournalTitle{Optics
  Express}}} \textbf{17}, 19662--19673 (2009).

\bibitem{Schroder2011}
S.~Schr{\"o}der, A.~Duparr{\'e}, L.~Coriand, \emph{et~al.}, \enquote{{Modeling
  of light scattering in different regimes of surface roughness},}
  \href{https://doi.org/10.1364/OE.19.009820}{{\protect\JournalTitle{Optics
  Express}}} \textbf{19}, 9820--9835 (2011).

\bibitem{Gunoven2020}
M.~G{\"u}n{\"o}ven, H.~Nasser, M.~{\"U}nal, \emph{et~al.},
  \enquote{{Implementation of generalized Harvey-Shack theory in light
  scattering from rough surfaces},}
  \href{https://doi.org/10.1103/PhysRevA.102.063521}{{\protect\JournalTitle{Physical
  Review A}}} \textbf{102}, 063521 (2020).

\bibitem{Zissa2018}
D.~E. Zissa, \enquote{{Diffracted radiance not proven to be fundamental
  quantity in nonparaxial scalar diffraction theory: comment},}
  \href{https://doi.org/10.1364/AO.57.004997}{{\protect\JournalTitle{Applied
  Optics}}} \textbf{57}, 4997 (2018).

\bibitem{Walther1968}
A.~Walther, \enquote{{Radiometry and Coherence},}
  \href{https://doi.org/10.1364/JOSA.58.001256}{{\protect\JournalTitle{Journal
  of the Optical Society of America}}} \textbf{58}, 1256--1259 (1968).

\bibitem{Wolf1978}
E.~Wolf, \enquote{{Coherence and radiometry},}
  \href{https://doi.org/10.1364/JOSA.68.000006}{{\protect\JournalTitle{Journal
  of the Optical Society of America}}} \textbf{68}, 6--17 (1978).

\bibitem{Nicodemus1963}
F.~E. Nicodemus, \enquote{{Radiance},}
  \href{https://doi.org/10.1119/1.1969512}{{\protect\JournalTitle{American
  Journal of Physics}}} \textbf{31}, 368--377 (1963).

\bibitem{CarterWolf1975}
W.~H. Carter and E.~Wolf, \enquote{{Coherence properties of lambertian and
  non-lambertian sources},}
  \href{https://doi.org/10.1364/JOSA.65.001067}{{\protect\JournalTitle{Journal
  of the Optical Society of America}}} \textbf{65}, 1067--1071 (1975).

\bibitem{CarterWolf1977}
W.~H. Carter and E.~Wolf, \enquote{{Coherence and radiometry with
  quasihomogeneous planar sources},}
  \href{https://doi.org/10.1364/JOSA.67.000785}{{\protect\JournalTitle{Journal
  of the Optical Society of America}}} \textbf{67}, 785--796 (1977).

\bibitem{Diffractsim2022}
R.~de~la Fuente~Herrezuelo, \enquote{{Diffractsim: A flexible Python
  diffraction simulator},}
  \href{https://doi.org/10.5281/zenodo.6147771}{Zenodo} (2022).
  \url{https://doi.org/10.5281/zenodo.6147771}.

\bibitem{KingmaBa2015}
D.~P. Kingma and J.~Ba, \enquote{{Adam}: A method for stochastic optimization,}
  in \href{https://arxiv.org/abs/1412.6980}{\emph{International Conference on
  Learning Representations,} } (2015).

\end{thebibliography}
}
  \let\bibliography\htmlbibliography
\fi
\bibliographystyle{opticajnl-threeauthors}
\title{Efficient convolution-based nonparaxial propagation of partially coherent light}
\author{Rafael de la Fuente Herrezuelo}
\email{rafael.fuente.herrezuelo@gmail.com}
\begin{abstract*}
We derive an efficient formulation for wide-angle propagation of partially coherent light from extended incoherent sources, extending the generalized van Cittert–Zernike Schell (GVS) propagator beyond the paraxial regime with absolute radiometric normalization. Within the Fresnel model, we establish that GVS and the classical paraxial source–PSF convolution are algebraically equivalent, explicitly deriving the corresponding source inversion, magnification, and absolute radiometric normalization. The resulting separation of mapped source and coherent pupil response enables wide-angle convolution in direction-cosine coordinates. Within a physically motivated diffracted-radiance formulation, this far-field convolution is exact for incoherent plane-wave mixtures. The same power prescription also relates delta-correlated source statistics to Lambertian radiance and hemispherical emitted power. At finite distances, we use a common incident curvature and approximate the source contributions by shifting and weighting a single coherent monitor response in direction-cosine coordinates. The differentiable formulation supports gradient-based inverse design of diffractive elements and metasurfaces represented by scalar pupils. Examples reveal geometric distortion and irradiance fall-off missed by the paraxial models and demonstrate wide-angle beam shaping at finite distances. A benchmark shows orders of magnitude faster evaluation than source-by-source propagation with a common error tolerance.
\end{abstract*}

\section{Introduction}
Diffraction under coherent illumination requires the propagation of one complex field. For extended sources such as LEDs, thermal emitters, and the Sun, distinct source points are often modeled as emitting independently, so their contributions add in intensity. After propagation, however, the fields at two receiving points can share contributions from the same emitters and become correlated. The van Cittert--Zernike relation connects these correlations to the source distribution \cite{VanCittert1934,Zernike1938}. It explains why spatially incoherent emission produces partially coherent illumination at a pupil and why source extent affects diffraction-fringe visibility.

The mutual intensity $J(\mathbf r_1,\mathbf r_2)=\langle U(\mathbf r_1)U^*(\mathbf r_2)\rangle$ describes this correlation \cite{Goodman1985,MandelWolf1965}. It depends on four transverse coordinates, although irradiance requires only its diagonal. Direct source integration, as in the Abbe description of partially coherent imaging, propagates each source contribution coherently and adds the intensities \cite{Hopkins1951,Mehta2014}. This accommodates nonparaxial propagation and changes in optical response, but its cost grows with the number of source samples. Random-field averaging also requires repeated propagations, with enough realizations to suppress statistical fluctuations. Both become costly when optical design requires many evaluations.

The illumination coherence provides a way to avoid many of these repeated propagations. Schell expressed the far-field intensity through the illumination coherence and pupil autocorrelation \cite{Schell1967}; Nugent extended the construction to the Fresnel region by including propagation curvature in the complex pupil \cite{Nugent1990}. Montoya et al. combined it with the source-to-pupil coherence relation in the generalized van Cittert--Zernike Schell (GVS) propagator \cite{Montoya2025}, enabling efficient evaluation through two-dimensional fast Fourier transforms (FFTs). Its derivation, however, uses paraxial source geometry. A more accurate coherent propagator does not by itself correct this source mapping.

At large angles, transverse slopes and direction cosines differ, so paraxial source mapping can misplace diffraction patterns and misrepresent their shape and strength. Harvey et al. formulated nonparaxial, coherent scalar diffraction in direction cosines, identifying the angular shift structure and its radiometric weights \cite{Harvey1999,Harvey2000Errata,HarveyVernold1998}. Nonparaxial transfer functions also describe oblique illumination in partially coherent imaging \cite{Muellers2022}. The challenge addressed here is to retain the computational advantage of one coherent response while incorporating wide-angle source geometry and a consistent absolute power scale.

These requirements matter in partially coherent holography \cite{Peng2021,Montoya2026,Sinha2026}, metasurface control of LED emission \cite{Poulsen2025,Khaidarov2020}, VCSEL-array beam shaping \cite{Ellingen2025}, and fan-out elements for structured-light 3D sensing \cite{Kim2020,Wu2024}. Inverse design repeatedly changes the pupil, making propagation cost important. It also requires more than the shape of a normalized image: the fraction of incident power delivered to the target determines collection efficiency. Fast propagation, wide-angle accuracy, and absolute normalization therefore need to be considered together.

We address these requirements by deriving a direction-cosine convolution that uses one coherent response per trial pupil and retains the FFT scaling of GVS. We first prove the equivalence between GVS and the classical source--PSF convolution within the Fresnel model, including source inversion, magnification, and absolute irradiance scaling. This new connection separates the mapped source from the coherent response and provides the basis for the nonparaxial extension. Under the physically motivated scalar power prescription we adopt, the normalization connects source covariance, Lambertian radiance, emitted power, and monitor irradiance, allowing collected power and collection efficiency to be calculated consistently. At finite distances, we assess separately the approximations for incident curvature and for shifting the monitor response in direction cosines.

A double slit tests fringe geometry and propagation accuracy, while a $3\times3$ phase grating reveals image distortion and irradiance fall-off missed by paraxial models. VCSEL-diffuser design tests wide-angle beam shaping, while a finite-distance $3\times3$ flat-top design improves spatial uniformity and spot shape over far-field designs with and without paraxial refocusing. Independent Rayleigh--Sommerfeld calculations assess accuracy. At 3,553 source samples, a USAF benchmark gives $164$--$174\times$ speedups over source-by-source angular-spectrum propagation at a common error tolerance. Sections~\ref{sec:sourceconv}--\ref{sec:angular} develop the formulation, Sections~\ref{sec:validation}--\ref{sec:results} present validation and applications, and Supplement~1 provides supporting derivations and numerical checks.

\section{From mutual intensity to the source--PSF convolution}
\label{sec:sourceconv}
\subsection{Geometry, source covariance, and notation}
\begin{figure}[!t]
\centering\includegraphics[width=\linewidth]{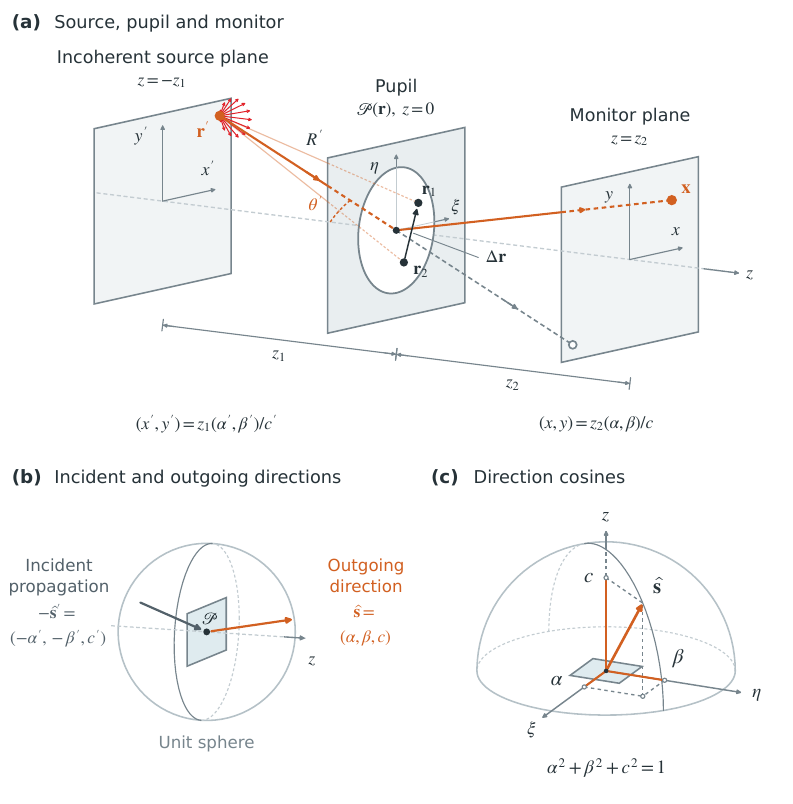}
\caption{Geometry and direction-cosine coordinates.
(a) A point $\mathbf{r}'=(x',y')$ in the incoherent source plane illuminates
two pupil points separated by $\Delta\mathbf{r}=\mathbf{r}_1-\mathbf{r}_2$.
Pupil and monitor coordinates are $\mathbf{r}=(\xi,\eta)$ and
$\mathbf{x}=(x,y)$, respectively. The dashed ray shows the incident chief
ray continuing without deflection. The red arrows show a Lambertian
angular intensity profile around the source normal.
(b) The unit sphere shows the incident and outgoing directions and the
central patch marks the pupil plane. The incident ray travels from the
source-side point $\widehat{\mathbf{s}}'=(\alpha',\beta',-c')$ to the origin,
so it propagates along $-\widehat{\mathbf{s}}'$. The outgoing ray leaves
the origin along $\widehat{\mathbf{s}}=(\alpha,\beta,c)$.
Their transverse propagation components differ by
$(\alpha+\alpha',\beta+\beta')$, where $c'=\cos\theta'$ and $c=\cos\theta$.
(c) The unit ray leaves the pupil patch. Its projections onto the axes give
$\alpha$, $\beta$, and $c$. Schematics are not to scale.}\label{fig:geometry}
\end{figure}

A source plane lies a distance $z_1$ before a thin pupil, followed by an observation plane at distance $z_2$ (Fig.~\ref{fig:geometry}). We use primed source coordinates $\mathbf r'=(x',y')$, pupil coordinates $\mathbf r=(\xi,\eta)$, and observation coordinates $\mathbf x=(x,y)$. The complex pupil transmission $\mathscr{P}(\mathbf r)$ describes the spatial response; an unqualified $P$ denotes radiant power. The scalar field is quasimonochromatic in a homogeneous, lossless medium. We use time dependence $e^{-i\omega t}$ and $k=2\pi/\lambda$.

For a complex field $U$, mutual intensity is $J(\mathbf r_1,\mathbf r_2)=\langle U(\mathbf r_1)U^*(\mathbf r_2)\rangle$, where brackets denote an ensemble average. An ideal spatially incoherent source is represented by
\begin{equation}
 J_0(\mathbf r'_1,\mathbf r'_2)
 =\lambda^2L_0(\mathbf r'_1)\delta^{(2)}(\mathbf r'_1-\mathbf r'_2).
 \label{eq:source}
\end{equation}
At this stage, $L_0$ is a nonnegative source weight, defined by the coefficient of the two-dimensional delta. Its radiometric interpretation is obtained in Section~\ref{sec:lambert}. The delta expresses independence of distinct emitters. 

\begin{table}[htbp]
\centering
\caption{Principal notation. Primed coordinates label the source; direction cosines are written explicitly.}
\label{tab:notation}
\begin{tabular}{@{}p{.28\linewidth}p{.67\linewidth}@{}}
\toprule
Symbol & Meaning\\
\midrule
$\mathbf r',\mathbf r,\mathbf x$ & Source, pupil, and observation-plane positions [m].\\
$J_0,J_1,J_b$ & Mutual intensity at the source, just before the pupil, and just after the pupil.\\
$L_0,\Gamma$ & Local source weight and the unnormalized, separation-dependent pupil coherence.\\
$Q,\operatorname{PSF}$ & Effective pupil including propagation phase, and its source-plane point-spread function.\\
$E_f(\mathbf x,z_1+z_2)$ & Detector irradiance, abbreviated $E(\mathbf x)$ [$\mathrm{W\,m^{-2}}$].\\
$(\alpha',\beta'),(\alpha,\beta)$ & Direction cosines towards the source and along the outgoing ray.\\
$\mathrm dP/\mathrm d\Omega,\ S$ & Radiant intensity and its cosine-divided angular output [$\mathrm{W\,sr^{-1}}$].\\
\bottomrule
\end{tabular}
\end{table}

Our two-dimensional Fourier convention is
\begin{equation}
 \widehat g(f_\xi,f_\eta)=\mathcal F\{g\}(f_\xi,f_\eta)
 =\iint g(\xi,\eta)e^{-i2\pi(f_\xi\xi+f_\eta\eta)}\,\mathrm d\xi\,\mathrm d\eta.
 \label{eq:conventions}
\end{equation}
Double integrals cover one transverse plane; quadruple integrals cover a pair of such planes. For example, $\mathrm d^2\mathbf r=\mathrm d\xi\,\mathrm d\eta$. Unspecified transverse domains are the full plane, and $\otimes$ denotes two-dimensional convolution. 

\subsection{Source coherence and the nested-transform expression}
Let $\Delta\mathbf r=\mathbf r_1-\mathbf r_2=(\Delta\xi,\Delta\eta)$. Fresnel propagation of the incoherent source gives the van Cittert--Zernike relation \cite{VanCittert1934,Zernike1938,Goodman1985},
\begin{align}
 J_1(\mathbf r_1,\mathbf r_2,z_1)
 &\simeq e^{i\pi(|\mathbf r_1|^2-|\mathbf r_2|^2)/(\lambda z_1)}\Gamma(\Delta\xi,\Delta\eta),\label{eq:vcz}\\
 \Gamma(\Delta\xi,\Delta\eta)
 &=\frac{1}{z_1^2}\iint L_0(x',y')
 e^{-i2\pi(\Delta\xi x'+\Delta\eta y')/(\lambda z_1)}\,\mathrm dx'\,\mathrm dy'.\label{eq:gamma}
\end{align}
The coherence factor $\Gamma$ retains its intensity scale: $\Gamma(0,0)$ is the pupil irradiance in this amplitude approximation. The quadratic phase accounts for the common incident wavefront curvature. Multiplication by the pupil gives $J_b=\mathscr{P}(\mathbf r_1)\mathscr{P}^*(\mathbf r_2)J_1$.

For Fresnel propagation to a detector at distance $z_2$, introduce the effective pupil
\begin{equation}
 Q(\xi,\eta)=\mathscr{P}(\xi,\eta)
 \exp\!\left[\frac{i\pi}{\lambda}\left(\frac{1}{z_1}+\frac{1}{z_2}\right)(\xi^2+\eta^2)\right].
 \label{eq:Qscreen}
\end{equation}
The aperture-autocorrelation reduction of Schell and Nugent \cite{Schell1967,Nugent1990}, combined with Eq.~\eqref{eq:gamma}, yields the structure of Eq.~\href{https://doi.org/10.1364/OE.571673}{(10)} in Montoya et al. \cite{Montoya2025}:
\begin{equation}
E(x,y)=\frac{1}{(\lambda z_2)^2}
\mathcal F\!\left\{
\Gamma(\Delta\xi,\Delta\eta)\,
\mathcal F^{-1}\!\left\{|\widehat Q|^2\right\}
(\Delta\xi,\Delta\eta)
\right\}\!\left(\frac{x}{\lambda z_2},\frac{y}{\lambda z_2}\right).
\label{eq:gvs}
\end{equation}
Here $E(x,y)=E_f(x,y,z_1+z_2)$, and equality is understood within the retained Fresnel model. The inverse transform of the coherent intensity $|\widehat Q|^2$ forms the pupil autocorrelation; multiplication by $\Gamma$ weights each pupil separation by its coherence; the outer transform gives detector irradiance. This two-dimensional sequence already avoids storing or propagating the four-coordinate mutual intensity. Supplement~1, Section~\ref{supp:proof}, derives it from the quadruple pupil-pair integral.

\subsection{Normalized equivalence of GVS and the source--PSF convolution}

We establish an explicit, absolutely normalized equivalence between the
GVS propagator and the classical paraxial source--PSF convolution
\cite{Goodman1985,Montoya2025} within the Fresnel model. The
product--convolution theorem identifies
$\widehat\Gamma\otimes|\widehat Q|^2$ in Eq.~\eqref{eq:gvs}. Fourier duality and coordinate
scaling applied to Eq.~\eqref{eq:gamma} give
\begin{equation}
 \widehat\Gamma(f_x,f_y)
 =\lambda^2L_0(-\lambda z_1f_x,-\lambda z_1f_y).
 \label{eq:gammahat}
\end{equation}
The two forward transforms produce source inversion, and the
two-dimensional scaling fixes $\lambda^2$. Transforming the integration
measure to detector coordinates and retaining the absolute normalization
of the source-plane PSF defined below gives
\begin{equation}
 E(x,y)=\frac{1}{M^2}\left[
 \lambda^2L_0\!\left(-\frac{x}{M},-\frac{y}{M}\right)
 \otimes\operatorname{PSF}(x,y)\right],\qquad
 M=\frac{z_2}{z_1}>0.
 \label{eq:psfconv}
\end{equation}
In the Fourier convolution, set $x'=-\lambda z_1\nu_x$ and
$y'=-\lambda z_1\nu_y$, so that
$\mathrm d\nu_x\,\mathrm d\nu_y
=(\lambda z_1)^{-2}\mathrm dx'\,\mathrm dy'$. Then
\begin{equation}
\begin{aligned}
E(x,y)
&=\frac{1}{\lambda^2z_1^2z_2^2}
  \iint L_0(x',y')
  \left|\widehat Q\!\left(
  \frac{x+Mx'}{\lambda z_2},
  \frac{y+My'}{\lambda z_2}
  \right)\right|^2
  \,\mathrm dx'\,\mathrm dy' \\
&=\iint\lambda^2L_0(x',y')
  \operatorname{PSF}(x+Mx',y+My')
  \,\mathrm dx'\,\mathrm dy'.
\end{aligned}
\label{eq:shiftedsum}
\end{equation}

Each emitter therefore contributes the same coherent intensity response,
translated by $(-Mx',-My')$. The area Jacobian $M^{-2}$ of
$u=-Mx'$, $v=-My'$ preserves the absolute irradiance scale.
This explicit separation of mapped source and coherent response provides
the starting point for the direction-cosine extension in
Section~\ref{sec:angular}. Supplement~1, Section~\ref{supp:proof}, gives the complete
transform and normalization proof.

Here the normalized source--PSF correspondence is explicit, and its
extension connects this coefficient to radiance, hemispherical exitance,
angular power, and monitor irradiance under the adopted power prescription
(Section~\ref{sec:lambert}).

For any coherent propagation algorithm, let $h(x,y;x',y')$ be the field response to a unit spatial delta at source position $(x',y')$. We define, in the Fresnel model:
\begin{equation}
\operatorname{PSF}(x,y)
=|h(x,y;0,0)|^2
=\frac{1}{\lambda^4z_1^2z_2^2}
 \left|\widehat Q\!\left(
 \frac{x}{\lambda z_2},
 \frac{y}{\lambda z_2}
 \right)\right|^2.
\label{eq:psf}
\end{equation}
The PSF retains the excitation's absolute normalization, without unit-peak or unit-integral rescaling. Propagating the central source field to the pupil, multiplying by $\mathscr{P}$, and propagating to the detector gives $h(x,y;0,0)$; lenses and phase elements enter the coherent model. The PSF's Fresnel expression in terms of $|\widehat Q|^2$, with all scale factors, is given in Supplement~1, Section~\ref{supp:proof}.

A source pitch $h_0$ maps to $h_s=Mh_0$, recovering Montoya et al.'s sampling relation \cite{Montoya2025}. The convolution grid must resolve both the mapped source and PSF. A Bluestein/chirp-$z$ transform \cite{Bluestein1970} permits a chosen output pitch and interval for an adequately sampled pupil. Different coherent
propagators can evaluate the response. One choice is the full scalar angular-spectrum propagator with the outgoing, decaying branch for evanescent components \cite{MatsushimaShimobaba2009}:
\begin{equation}
\operatorname{PSF}(x,y)= \left|\mathcal P_{z_2}\{U_{\mathscr{P}}\}(x,y)\right|^2 =
\left|
\mathcal F^{-1}\!\left\{
\widehat U_{\mathscr{P}}(f_x,f_y)\,
e^{ikz_2\sqrt{1-\lambda^2(f_x^2+f_y^2)}}
\right\}(x,y)
\right|^2.
\label{eq:ASM}
\end{equation}
Here we define $U_{\mathscr{P}}(\xi, \eta)=\frac{e^{i k z_1}}{i \lambda z_1} \mathscr{P}(\xi,\eta)
e^{i\pi(\xi^2+\eta^2)/(\lambda z_1)}$. To reuse one response for the whole source, the response to each emitter must be a copy of the central intensity pattern shifted
by $(-Mx',-My')$. This holds in the Fresnel model and gives
Eq.~\eqref{eq:shiftedsum}. Calculating the central response with
a more accurate propagator does not, by itself, ensure that the
other responses remain shifted copies. Section~\ref{sec:angular}
shows how to reuse a response in direction cosines and explains
the approximations needed at finite distances.

\FloatBarrier
\section{Nonparaxial convolution extension and its radiometric consequences}
\label{sec:angular}
The nonparaxial construction follows the same sequence: express the illumination coherence, form one coherent pupil response, and combine the two by convolution. True direction cosines replace the small-angle coordinate map. The angular power relation first establishes the output quantity; its application to the source covariance then gives the physical meaning of $L_0$.

\subsection{From pupil coherence to angular power}
\label{sec:angular_power}
An outgoing ray has direction cosines $(\alpha,\beta,c)$, with $c=\cos\theta=\sqrt{1-\alpha^2-\beta^2}>0$. The angular domain is $\mathcal D=\{(\alpha,\beta):\alpha^2+\beta^2<1\}$. We adopt the coherent angular power relation of Harvey et al.'s scalar diffracted-radiance formulation, including its erratum \cite{Harvey1999,Harvey2000Errata}. In this relation, the squared coherent Fourier amplitude, with its wavelength scaling, represents area-integrated diffracted radiance, and multiplication by $c$ gives radiant intensity. We extend the relation to partially coherent fields by ensemble averaging:
\begin{align}
 \frac{\mathrm dP}{\mathrm d\Omega}(\alpha,\beta)
 &=\frac{c}{\lambda^2}\iiiint J_b(\mathbf r_1,\mathbf r_2)\nonumber\\
 &\qquad\times e^{-ik[\alpha(\xi_1-\xi_2)+\beta(\eta_1-\eta_2)]}
 \,\mathrm d^2\mathbf r_1\,\mathrm d^2\mathbf r_2.
 \label{eq:angularoutput}
\end{align}
The exponential resolves the outgoing direction, and the cosine is the projected-area factor converting area-integrated diffracted radiance to power per solid angle. The four integration coordinates are the two components of each pupil point. The outgoing scalar far-field approximation requires negligible phase from omitted quadratic path terms. Comparisons of Harvey--Shack predictions with measured angular scattering distributions provide empirical motivation for this scalar framework \cite{Harvey1999,Schroder2011,Gunoven2020}. Equation \eqref{eq:angularoutput} combines the far-field Rayleigh–Sommerfeld kernel with the adopted plane-flux identification. Supplement 1, Section \ref{supp:angular_power}, discusses this identification, its use in Harvey et al.’s erratum, and the question raised by Zissa \cite{Zissa2018}. The convolution and covariance–radiance correspondence below follow from this stated radiometric input.

A vector-aperture benchmark also motivates this choice: for a prescribed scalar pupil under normally incident unpolarized illumination, the single-cosine factor is the geometric mean of the two polarization-dependent obliquity factors and approximates their arithmetic mean, $(1+\cos^2\theta)/2$, with an error of $O(\theta^4)$ (Supplement~1, Section~\ref{supp:angular_power}).

For computation, define the area-integrated angular output
\begin{equation}
 S(\alpha,\beta)=\frac{1}{c}\frac{\mathrm dP}{\mathrm d\Omega}(\alpha,\beta),
 \qquad P_{\mathcal R}=\iint_{\mathcal R}S(\alpha,\beta)\,\mathrm d\alpha\,\mathrm d\beta.
 \label{eq:radiometry}
\end{equation}
The power relation follows from $\mathrm d\alpha\,\mathrm d\beta=c\,\mathrm d\Omega$, with $\mathrm d\Omega=\sin\theta\,\mathrm d\theta\,\mathrm d\varphi$. Both angular quantities are reported in $\mathrm{W\,sr^{-1}}$, with their integration measures stated explicitly. A distant plane detector receives $E_{\mathrm{det}}=c^3(\mathrm dP/\mathrm d\Omega)/z_2^2$ at $(x,y)=z_2(\alpha,\beta)/c$. This converts the angular result to detector irradiance.

\subsection{True source directions and the geometric cancellation}
\label{sec:source_directions}
For a source point $(x',y',-z_1)$, let
\begin{equation}
 R'=\sqrt{z_1^2+x'^2+y'^2},\qquad
 \alpha'=\frac{x'}{R'},\quad\beta'=\frac{y'}{R'},\quad
 c'=\frac{z_1}{R'}=\sqrt{1-\alpha'^2-\beta'^2}.
 \label{eq:sourceangles}
\end{equation}
The primed direction points towards the emitter, so the arriving wave has transverse direction $(-\alpha',-\beta')$. These are exact geometrical coordinates. The source-to-angle map is
\begin{equation}
 x'=\frac{z_1\alpha'}{c'},\qquad y'=\frac{z_1\beta'}{c'},\qquad
 L_{\mathrm{in}}(\alpha',\beta')=L_0\!\left(\frac{z_1\alpha'}{c'},\frac{z_1\beta'}{c'}\right).
 \label{eq:mappedradiance}
\end{equation}
Here $L_{\mathrm{in}}$ is the angular version of the source weight. Evaluating the incident propagation amplitude at the pupil center gives the squared weight $z_1^2/R'^4$. The Jacobian of the exact coordinate map is $z_1^2/c'^4$; their product is unity. Consequently, the stationary incident coherence is
\begin{equation}
 \Gamma(\Delta\xi,\Delta\eta)
 =\iint_{\mathcal D}L_{\mathrm{in}}(\alpha',\beta')
 e^{-ik(\alpha'\Delta\xi+\beta'\Delta\eta)}\,\mathrm d\alpha'\,\mathrm d\beta'.
 \label{eq:angularcoherence}
\end{equation}
Supplement~1, Section~\ref{supp:angular}, gives the cancellation explicitly. For a prescribed incoherent mixture of plane waves, Eq.~\eqref{eq:angularcoherence} is exact within the scalar model. A finite source requires control of the pupil curvature and amplitude variations; Section~\ref{sec:curvature} states the approximation retained by the implementation.

\begin{figure}[!htbp]
\centering\includegraphics[width=\linewidth]{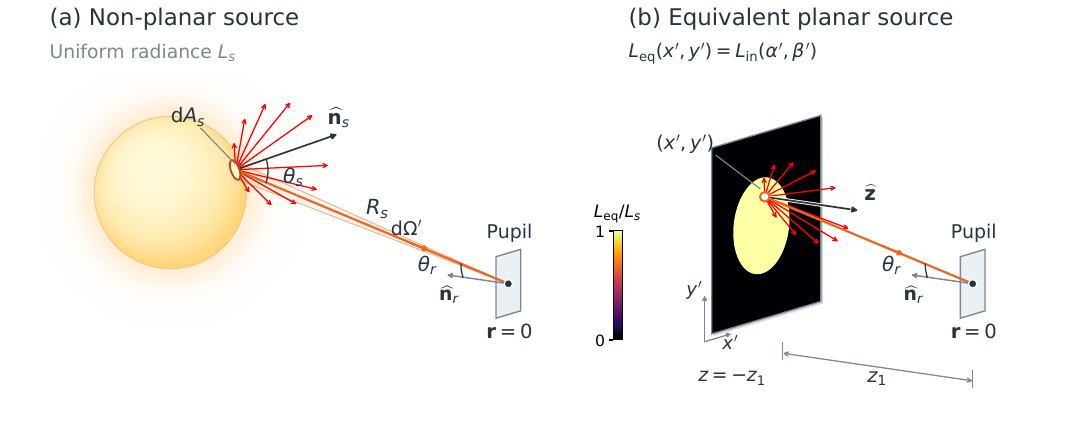}
\caption{Radiometric mapping of a nonplanar emitter.
(a) A visible surface element contributes radiance along
the direction connecting it to the pupil center.
(b) An equivalent planar distribution reproduces the
same directional radiance at that point. Finite-distance
conditions are discussed in Supplement~1, Section~\ref{supp:incident}.}
\label{fig:nonplanar_source}
\end{figure}

For quasihomogeneous sources in the radiometric limit, radiance
is conserved along rays in a homogeneous, lossless
medium~\cite{Walther1968,Wolf1978,Nicodemus1963}.
We can therefore construct $L_{\mathrm{in}}$ for nonplanar
emitters by tracing rays from the pupil center to the visible
emitting surface (Fig.~\ref{fig:nonplanar_source});
Eq.~\eqref{eq:mappedradiance} is the parallel-plane case. At finite distances,
using one coherent response also requires control of
variations across the pupil and
direction-dependent curvature (Supplement~1, Section~\ref{supp:incident}).

\subsection{Angular convolution and fast numerical evaluation}
For stationary illumination, $J_1=\Gamma(\Delta\mathbf r)$. Define the coherent angular pupil response
\begin{equation}
 H_\infty(\alpha,\beta)=\frac1{\lambda^2}
 \left|\mathcal F\{\mathscr{P}\}\!\left(\frac\alpha\lambda,\frac\beta\lambda\right)\right|^2.
 \label{eq:Hinf}
\end{equation}
Substituting Eq.~\eqref{eq:angularcoherence} into Eq.~\eqref{eq:angularoutput} gives
\begin{equation}
 \begin{aligned}
 S(\alpha,\beta)
 &=\iint_{\mathcal D}L_{\mathrm{in}}(\alpha',\beta')
 H_\infty(\alpha+\alpha',\beta+\beta')\,\mathrm d\alpha'\,\mathrm d\beta'\\
 &=L_{\mathrm{in}}(-\alpha,-\beta)\otimes H_\infty(\alpha,\beta),\qquad
 \frac{\mathrm dP}{\mathrm d\Omega}=\cos\theta\,S.
 \end{aligned}
 \label{eq:angularconv}
\end{equation}
Every incident direction shifts the same coherent diffraction pattern. The plus signs follow from the primed source-direction convention and produce the inverted source map inside the convolution. Direction-cosine shift invariance supplies the underlying angular structure \cite{Harvey1999,HarveyVernold1998}. Furthermore, an angle-dependent pupil that changes the transmission with a weight $t(\alpha',\beta')$ that is uniform across the pupil can be absorbed into the incident radiance. The pupil’s angular dependence factors as $\mathscr{P}\left(\xi, \eta ; \alpha^{\prime}, \beta^{\prime}\right)=t\left(\alpha^{\prime}, \beta^{\prime}\right) \mathscr{P}_0(\xi, \eta)$. The same convolution then simply uses the effective illumination $L_{\text {eff }}\left(\alpha^{\prime}, \beta^{\prime}\right)=\left|t\left(\alpha^{\prime}, \beta^{\prime}\right)\right|^2 L_{\text {in }}\left(\alpha^{\prime}, \beta^{\prime}\right)$.

The calculation uses one coherent pupil response and a two-dimensional FFT convolution. The response and its transform can be cached while illumination changes. Evaluating $L_0$ at the coordinates in Eq.~\eqref{eq:mappedradiance} incorporates the source geometry. The full transverse-frequency response is retained during convolution; the physical outgoing disk is selected afterwards, without renormalization. This preserves all frequency shifts needed for allowed incident and outgoing directions.

\subsection{From source covariance to Lambertian radiance and exitance}
\label{sec:lambert}
Equation~\eqref{eq:source} specifies a delta-correlated covariance with nonnegative weight $L_0$; no angular emission law has been prescribed. Applying the adopted angular power relation, Eq.~\eqref{eq:angularoutput}, directly to this covariance determines both the emission law and the radiometric meaning of the source weight. The delta sets the two source positions equal, so the Fourier phase vanishes and the wavelength factors cancel:
\begin{equation}
 \frac{\mathrm dP_0}{\mathrm d\Omega}(\theta,\varphi)
 =\cos\theta\iint L_0(x',y')\,\mathrm dx'\,\mathrm dy'.
 \label{eq:lambert}
\end{equation}
An elementary source area $\mathrm dA'$ therefore contributes $L_0\cos\theta\,\mathrm dA'$ to radiant intensity. By the definition of radiance,
\begin{equation}
 L(x',y';\theta,\varphi)
 =\frac{\mathrm d^2P_0}{\mathrm dA'\cos\theta\,\mathrm d\Omega}
 =L_0(x',y').
 \label{eq:radiance_identification}
\end{equation}
The radiance is independent of emission direction, while radiant intensity decreases with the projected source area. Thus the derivation reveals Lambertian emission within the adopted scalar formulation. Its radiant exitance, the power emitted per unit source area into the forward hemisphere $\Omega_+$, is
\begin{equation}
 I_0(x',y')=\iint_{\Omega_+}L_0(x',y')\cos\theta\,\mathrm d\Omega
 =\pi L_0(x',y').
 \label{eq:exitance}
\end{equation}
Equations~\eqref{eq:lambert}--\eqref{eq:exitance} establish the explicit correspondence between delta-correlated source statistics, angle-independent radiance, and hemispherical exitance. In particular, they identify the exitance-normalized covariance coefficient as $(\lambda^2/\pi)I_0$, connecting the angular derivation to the normalization used by Goodman and GVS \cite{Goodman1985,Montoya2025}. Supplement~1, Section~\ref{supp:carterwolf}, compares this result with Carter and Wolf's sinc-correlated Lambertian source under their ray-normal power prescription \cite{CarterWolf1975}; equal emitted radiance in the homogeneous-source limit does not imply equal covariances or equal finite-pupil predictions.

The geometric cancellation in Section~\ref{sec:source_directions} now has a corresponding radiometric meaning: $L_{\mathrm{in}}$ is source radiance carried along each ray, and $\mathrm d\alpha'\,\mathrm d\beta'=c'\,\mathrm d\Omega'$ supplies the receiving-plane projection. Evaluating Eq.~\eqref{eq:angularcoherence} at zero separation gives the pupil irradiance, $\Gamma(0,0)=\iint L_{\mathrm{in}}c'\,\mathrm d\Omega'$. The coherence calculation and the radiometric description consequently share the same absolute scale \cite{CarterWolf1977,Wolf1978}.

Localizing $L_0$ at fixed emitted power produces a point emitter that retains the cosine emission pattern. Selecting one arrival direction in $L_{\mathrm{in}}$ instead produces a plane wave and its shifted coherent diffraction pattern. Supplement~1, Section~\ref{supp:angular}, gives the coefficients and measures for these spatial and angular limits. Section~\ref{supp:checks} verifies the source normalization and point-emitter limit; Section~\ref{supp:convergence} tests extended angular sources at infinity.

\subsection{Finite source distance: one common pupil curvature}
\label{sec:curvature}
A nearby emitter illuminates the pupil with a curved wavefront, and the curvature varies with its direction. To preserve a single reusable response, the finite-distance implementation retains the exact source-direction map and uses the common on-axis quadratic phase
\begin{equation}
\boxed{\begin{gathered}
 Q_{z_1}(\mathbf r)=\mathscr{P}(\mathbf r)e^{i\pi|\mathbf r|^2/(\lambda z_1)},\qquad
 H_{z_1}(\alpha,\beta)=\frac1{\lambda^2}
 \left|\mathcal F\{Q_{z_1}\}\!\left(\frac\alpha\lambda,\frac\beta\lambda\right)\right|^2,\\
 S(\alpha,\beta)\simeq L_{\mathrm{in}}(-\alpha,-\beta)\otimes H_{z_1}(\alpha,\beta),\qquad
 \frac{\mathrm dP}{\mathrm d\Omega}=\cos\theta\,S.
\end{gathered}}
 \label{eq:common}
\end{equation}
This approximation assigns all source points the same quadratic curvature and evaluates their propagation amplitudes at the pupil center. Its output is the outgoing angular far field.

For a pupil enclosed by radius $a$, define $F_1=a^2/(\lambda z_1)$. At central incidence $\theta_0$, the largest leading quadratic-phase mismatch over that enclosing disk is
\begin{equation}
 \chi_0=\pi F_1\bigl[1-\cos^{3}\theta_0\bigr].
 \label{eq:phasecriterion}
\end{equation}
The diagnostic $\chi_0$ combines Fresnel number and incidence angle: smaller $F_1$ reduces the curvature mismatch, while larger $\theta_0$ increases its anisotropy. It is a phase mismatch, not an intensity-error bound; source width, pupil weighting, amplitude variation, and higher-order path terms also matter. Residual error can remain on axis where $\chi_0=0$. Supplement~1, Section~\ref{supp:central_response}, derives the diagnostic; Figs.~\ref{fig:profiles} and \ref{fig:maps} show the fringe-shape changes and error trends.

For the double slit at $F_1=0.8$ and $\theta_0=40^\circ$, matching the central emitter's quadratic curvature reduces the error from 12.5\% to 3.86\%; matching its full phase and amplitude reduces it to 0.185\% (Supplement~1, Section~\ref{supp:central_response}). Both retain one response and approximate variation across the extended source. The maps below use common on-axis curvature.

\subsection{Finite monitor distance}
\label{sec:monitor}
For a monitor at finite distance $z_2>0$, define the coherent monitor-plane response $H^{\mathrm{mon}}_{z_1,z_2}(x,y)=\left(\lambda z_1\right)^2 \operatorname{PSF}(x, y)=|\mathcal P_{z_2}\{Q_{z_1}\}(x,y)|^2$, the intensity response to the reference pupil in Eq.~\eqref{eq:common}, normalized by the incident reference irradiance. It is not normalized to unit integral. Here $\mathcal P_{z_2}$ is the full angular-spectrum propagator of Eq.~\eqref{eq:ASM}, equivalently the exact scalar Rayleigh--Sommerfeld propagator. Write $\rho(x,y)=\sqrt{z_2^2+x^2+y^2}$ and $c=c(\alpha,\beta)=\sqrt{1-\alpha^2-\beta^2}$. Mapping the reference response to direction cosines, convolving with the angular illumination, and mapping back to the monitor gives
\begingroup\setlength{\fboxsep}{3pt}
\begin{equation}
\boxed{\begin{aligned}
E(x,y;z_1,z_2)
&\simeq\frac{z_2^2}{\rho^4(x,y)}
\Bigg[L_{\mathrm{in}}(-\alpha,-\beta)\otimes
\Bigg[\frac{z_2^2}{c^4}H^{\mathrm{mon}}_{z_1,z_2}\!\left(
\frac{z_2\alpha}{c},\frac{z_2\beta}{c}\right)\Bigg]
\Bigg]_{(\alpha,\beta)=\frac{(x,y)}{\rho(x,y)}}\\[1ex]
&=\iint_{\mathcal D}L_{\mathrm{in}}(\alpha',\beta')
\left(\frac{c}{c_s}\right)^4 H^{\mathrm{mon}}_{z_1,z_2}\!\left(
\frac{z_2(\alpha+\alpha')}{c_s},\frac{z_2(\beta+\beta')}{c_s}\right)
\,\mathrm d\alpha'\,\mathrm d\beta'.
\end{aligned}}
\label{eq:monitorconv}
\end{equation}
\endgroup
The convolution acts over direction cosines. In the integral, $(\alpha,\beta)=(x,y)/\rho(x,y)$, $c=z_2/\rho(x,y)$, and $c_s=c(\alpha+\alpha',\beta+\beta')$. For shifted arguments outside the unit disk, the response in the inner brackets is continued by $H_{z_1}$ from Eq.~\eqref{eq:common}; equivalently, the weighted monitor response in the integral is replaced by $(c^4/z_2^2)H_{z_1}(\alpha+\alpha',\beta+\beta')$ when $(\alpha+\alpha')^2+(\beta+\beta')^2\ge1$.

For parallel source and monitor planes, the geometric-image limit carries the $\cos^4\theta$ weight on $L_0(-x/M,-y/M)$, with $M=z_2/z_1$. Equation~\eqref{eq:monitorconv} also retains the angular remapping of the diffraction response.

Equation~\eqref{eq:monitorconv} recovers the angular convolution, Eq.~\eqref{eq:common}, in the outgoing far-field limit and the GVS convolution, Eqs.~\eqref{eq:psfconv}--\eqref{eq:shiftedsum}, equivalently the nested-transform Eq.~\eqref{eq:gvs}, in the Fresnel/paraxial limit. It reproduces the axial incident component at any monitor distance for the prescribed $Q_{z_1}$. At finite distance, remapping introduces an outgoing-propagation approximation, distinct from the common incident curvature at finite $z_1$ (Supplement~1, Section~\ref{supp:monitor_remap}). For a finite source plane, the full Rayleigh--Sommerfeld source integral retains both propagation legs:
\begin{equation}
\begin{aligned}
E(x,y;z_1,z_2)=\iint_{\mathcal D} &L_{\mathrm{in}}(\alpha',\beta')
\Biggl|\iint_{\mathbb R^2}\mathscr{P}(\xi,\eta)
 \frac{R'^2}{R_1^2}\left(1+\frac{i}{kR_1}\right)e^{ik(R_1-R')}\\[-2pt]
&\quad\times\frac{z_2}{i\lambda R^2}\left(1+\frac{i}{kR}\right)
 e^{ikR}\,\mathrm d\xi\,\mathrm d\eta\Biggr|^2
 \,\mathrm d\alpha'\,\mathrm d\beta'.
\end{aligned}
\label{eq:monitordirect}
\end{equation}
Here $R=\sqrt{z_2^2+(x-\xi)^2+(y-\eta)^2}$, $R_1=\sqrt{z_1^2+(\xi-x')^2+(\eta-y')^2}$, $R'=z_1/c'$, and $(x',y')=z_1(\alpha',\beta')/c'$ as in Eq.~\eqref{eq:mappedradiance}. The factor $R'^2/R_1^2$ follows from the source covariance in Eq.~\eqref{eq:source} and the source-to-angle Jacobian; the source-dependent constant phase $e^{ikR'}$ cancels in the intensity. This Abbe or sum-over-source calculation \cite{Hopkins1951,Mehta2014} retains each emitter's incident amplitude and wavefront, as well as the full outgoing RS kernel, whereas Eq.~\eqref{eq:monitorconv} reuses one response. Table~\ref{tab:scope} distinguishes their scopes.

\begin{table}[!htbp]
\centering
\caption{Model scope and checks, including separable angular transmission.}
\label{tab:scope}
\begin{tabular}{@{}>{\raggedright\arraybackslash}p{.27\linewidth}>{\raggedright\arraybackslash}p{.68\linewidth}@{}}
\toprule
Formulation & Scope and required check\\
\midrule
Angular convolution,\newline Eq.~\eqref{eq:angularconv} & Exact for a prescribed incoherent plane-wave mixture under the adopted angular power relation. Check the outgoing far-field condition.\\[3pt]
Finite source distance,\newline Eq.~\eqref{eq:common}& Approximates a finite source by a common incident phase and pupil-center amplitudes. Compare with Eq.~\eqref{eq:monitordirect} to assess their combined error.\\[3pt]
Finite source and monitor distance,\newline Eq.~\eqref{eq:monitorconv}& Combines the incident approximation with outgoing-response remapping. Compare with Eq.~\eqref{eq:monitordirect} to assess their combined error.\\[3pt]
RS source integral,\newline Eq.~\eqref{eq:monitordirect} & Full scalar propagation from each source point through the pupil to the monitor, retaining both RS kernels.\\
\bottomrule
\end{tabular}
\end{table}

\subsection{Diffractsim implementation}
\label{sec:implementation}

The Diffractsim extension \cite{Diffractsim2022} accepts a coherent
monitor PSF or angular pupil response. Full angular-spectrum propagation
or another coherent engine can supply it, with coordinates, sample weights,
and absolute normalization consistent with the formulation in
Table~\ref{tab:scope}.

At fixed pupil, geometry, wavelength, and numerical grids, the coherent
intensity response and its Fourier transform can be cached while the
illumination changes. Conversely, the mapped-source transform can be
cached while the pupil varies. Writing the convolution as
$E=G\otimes\operatorname{PSF}$, where $G$ includes the source mapping and
absolute normalization, differentiation with respect to a pupil
parameter $p$ that leaves $G$ fixed gives
\begin{equation*}
 \frac{\partial E}{\partial p}
 =G\otimes\frac{\partial\operatorname{PSF}}{\partial p}.
\end{equation*}
The angular convolution has the same derivative structure. This
separates illumination averaging from coherent-response differentiation
and supports repeated forward and gradient calculations while retaining
the FFT scaling of GVS. An uncached calculation uses the same four
transforms as GVS when the coherent response requires one transform.

\section{Validation strategy}
\label{sec:validation}
We first test that equivalent formulas give the same result and use the same power scale. Three separately coded calculations agree to floating-point precision when they use identical integration weights. Hemispherical integration and the transparent-plane limit check the normalization (Supplement~1, Sections~\ref{supp:angular}--\ref{supp:checks}).

We then test the propagation accuracy against RS calculations for the same illumination and pupil. The angular reference uses each emitter's full RS field at the pupil and adds their far-field intensities. Both predictions use the same angular power relation. For a finite monitor, Eq.~\eqref{eq:monitordirect} includes propagation from source to pupil and from pupil to monitor. The collimated VCSEL illumination is specified at the diffuser, so only the outgoing propagation is needed.

Writing $R(\alpha,\beta)\equiv\mathrm dP/\mathrm d\Omega=cS$, we compare angular distributions over a common region $\mathcal R$ using
\begin{equation}
\epsilon_R=
\left[
\frac{\iint_{\mathcal R}|R-R_{\mathrm{ref}}|^2\,\mathrm d\Omega}
{\iint_{\mathcal R}|R_{\mathrm{ref}}|^2\,\mathrm d\Omega}
\right]^{1/2}
=
\left[
\frac{\iint_{\mathcal R}c|S-S_{\mathrm{ref}}|^2\,\mathrm d\alpha\,\mathrm d\beta}
{\iint_{\mathcal R}c|S_{\mathrm{ref}}|^2\,\mathrm d\alpha\,\mathrm d\beta}
\right]^{1/2}.
\label{eq:error}
\end{equation}
Both predictions retain their absolute scale. We do not fit an amplitude or shift either distribution. Monitor errors use the corresponding relative $L^2$ norm over detector area. The comparison regions are specified in Supplement~1, Section~\ref{supp:checks}. Display normalization does not affect these errors. Shape error and spatial nonuniformity are separate measures of design quality; their definitions and normalizations are given with each example.

To check numerical accuracy, we refine source and pupil integration, numerical grids, interpolation, and comparison regions separately. In three representative cases from the validity map, the largest refinement change is below one tenth of the reported model error (Supplement~1, Section~\ref{supp:convergence}). These checks establish that the observed discrepancies mainly come from the propagation approximations.

Timing comparisons use the fastest tested settings that meet the same irradiance-error tolerance. Each timed calculation includes the optical setup and detector interpolation; Eq.~\eqref{eq:monitorconv} rebuilds its coherent response on every call (Supplement~1, Section~\ref{supp:timing}).

\section{Results}
\label{sec:results}
Using these checks, we test when nonparaxial propagation improves
diffraction predictions and optical design. We compare the paraxial
model, Eq.~\eqref{eq:psfconv}, with the angular far-field and
finite-monitor models, Eqs.~\eqref{eq:common} and
\eqref{eq:monitorconv}. The independent RS calculation,
Eq.~\eqref{eq:monitordirect}, checks both their predictions and
the resulting designs. We then compare computation times at
a common error tolerance.

\subsection{Double slit: geometry and finite-distance accuracy}
We begin with a double slit because its fringes reveal errors in
position, spacing, and visibility. At $\lambda=1~\mu$m, the slits
are $3.125~\mu$m wide and $7.125~\mu$m high, with centers
$12~\mu$m apart and enclosing radius $a=8.3596~\mu$m.
We describe the Gaussian incoherent source using slope coordinates
$\mathbf t=\mathbf r'/z_1$:
\begin{equation}
 L_0(z_1\mathbf t)=\frac{L_*}{2\pi\sigma_t^2}\exp\!\left[-\frac{|\mathbf t-\mathbf t_0|^2}{2\sigma_t^2}\right],\qquad
 \mathbf t_0=(\tan\theta_0,0),
 \label{eq:gaussian}
\end{equation}
where $L_*=1~\mathrm{W\,m^{-2}\,sr^{-1}}$ and $\sigma_t=0.025$. Keeping this distribution fixed separates changes in source distance from changes in central incidence or source angular extent. Here $F_1=a^2/(\lambda z_1)$ is the incoming Fresnel number; $F_2=a^2/(\lambda z_2)$ describes the monitor distance.

\begin{figure}[!htbp]
\centering\includegraphics[width=\linewidth]{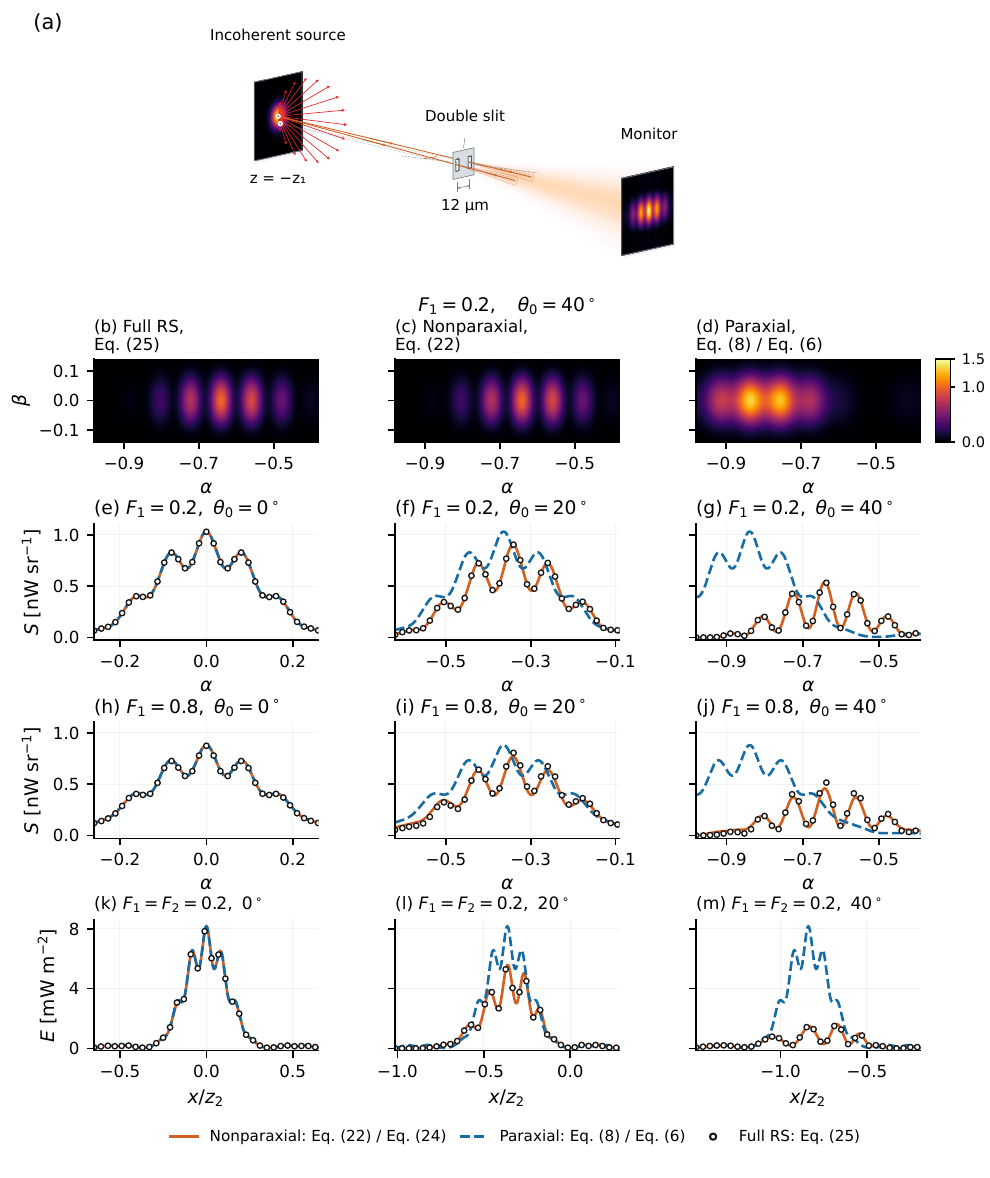}
\caption{Double-slit geometry and diffraction. (a) Gaussian incoherent source, pupil, and output geometry. (b--d) Angular radiant intensity at $F_1=0.2$, $\theta_0=40^\circ$, on a common scale divided by the RS maximum; the displayed $\beta$ range is cropped. (e--j) Absolute angular-density profiles at $\beta=0$ for $F_1=0.2$ and $0.8$. (k--m) Flat-monitor irradiance at $y=0$, $F_1=F_2=0.2$, plotted against physical slope $x/z_2$. In (e--m), the incidence angles are $0^\circ$, $20^\circ$, and $40^\circ$ from left to right. Angular panels use Eqs.~\eqref{eq:psfconv}, \eqref{eq:common}, and Supplement~1, Eq.~\eqref{eq:fullreference}, the far-field limit of Eq.~\eqref{eq:monitordirect}. The bottom row uses Eqs.~\eqref{eq:psfconv}, \eqref{eq:monitorconv}, and the full RS integral, Eq.~\eqref{eq:monitordirect}.}
\label{fig:profiles}
\end{figure}

Figure~\ref{fig:profiles}(b--j) first tests the angular predictions against full source-to-pupil RS propagation. At oblique incidence, identifying slope with direction cosine displaces the paraxial pattern and changes its amplitude and fringe scale. Equation~\eqref{eq:common} reproduces the RS fringe positions, spacing, and absolute amplitudes much more closely than the paraxial model [Fig.~\ref{fig:profiles}(b--j)]. The paraxial approximation identifies source slope with direction cosine, shifting the pattern and misrepresenting its fringe structure and strength. At $F_1=0.2$ and $\theta_0=40^\circ$, Eq.~\eqref{eq:common} reduces the relative angular error from approximately $186\%$ for Eq.~\eqref{eq:psfconv} to $0.83\%$. The improvement also appears on a finite monitor [Fig.~\ref{fig:profiles}(k--m)], with $F_1=F_2=0.2$. Equation~\eqref{eq:monitorconv} captures the broadening and fall in peak irradiance with incidence angle that the paraxial translation misses. Both models use the same axial coherent response, so this comparison tests how they combine that response with the extended illumination. Against the full RS reference, Eq.~\eqref{eq:monitorconv} remains below $1.8\%$ relative $L^2$ image error at all three tested angles.

\begin{figure}[!htbp]
\centering\includegraphics[width=\linewidth]{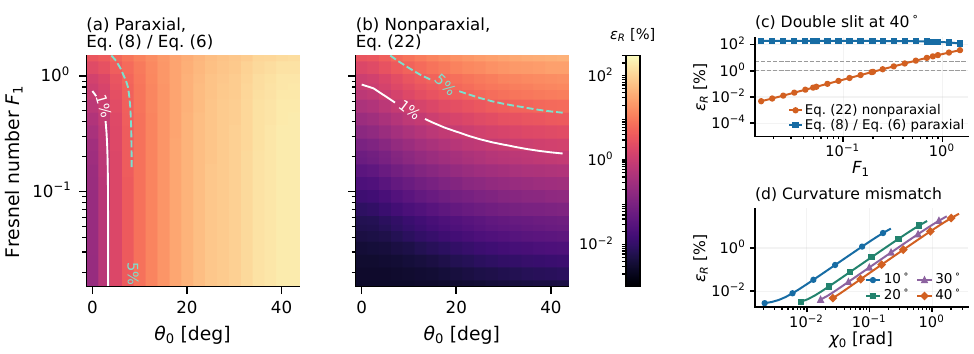}
\caption{Double-slit validity study (396 cases; $F_1=a^2/(\lambda z_1)$). Absolute-scale radiant-intensity errors for (a) equivalent paraxial Eqs.~\eqref{eq:psfconv}/\eqref{eq:gvs} and (b) nonparaxial Eq.~\eqref{eq:common}; contours mark $1\%$ and $5\%$. (c) Errors versus $F_1$ at $40^\circ$. (d) Nonparaxial error versus leading curvature mismatch for four incidence angles, with $F_1$ varying along each curve. }
\label{fig:maps}
\end{figure}

The 396-case sweep confirms that Eq.~\eqref{eq:common} extends the regions below $1\%$ and $5\%$ error [Fig.~\ref{fig:maps}(a--c)]. At $40^\circ$ and $F_1=0.8$, it reduces the error from approximately $165\%$ to $12.5\%$. The remaining error reflects the mismatch between oblique wavefront curvature and the common on-axis phase, whose trend is tracked by $\chi_0$ [Fig.~\ref{fig:maps}(d)]. Section~\ref{sec:curvature} describes how matching the central emitter improves this response; Supplement~1, Section~\ref{supp:convergence}, provides the numerical checks.

\FloatBarrier
\subsection{Finite-monitor imaging with a $3\times3$ phase grating}
\label{sec:grating}
\begin{figure}[!t]
\centering\includegraphics[width=\linewidth]{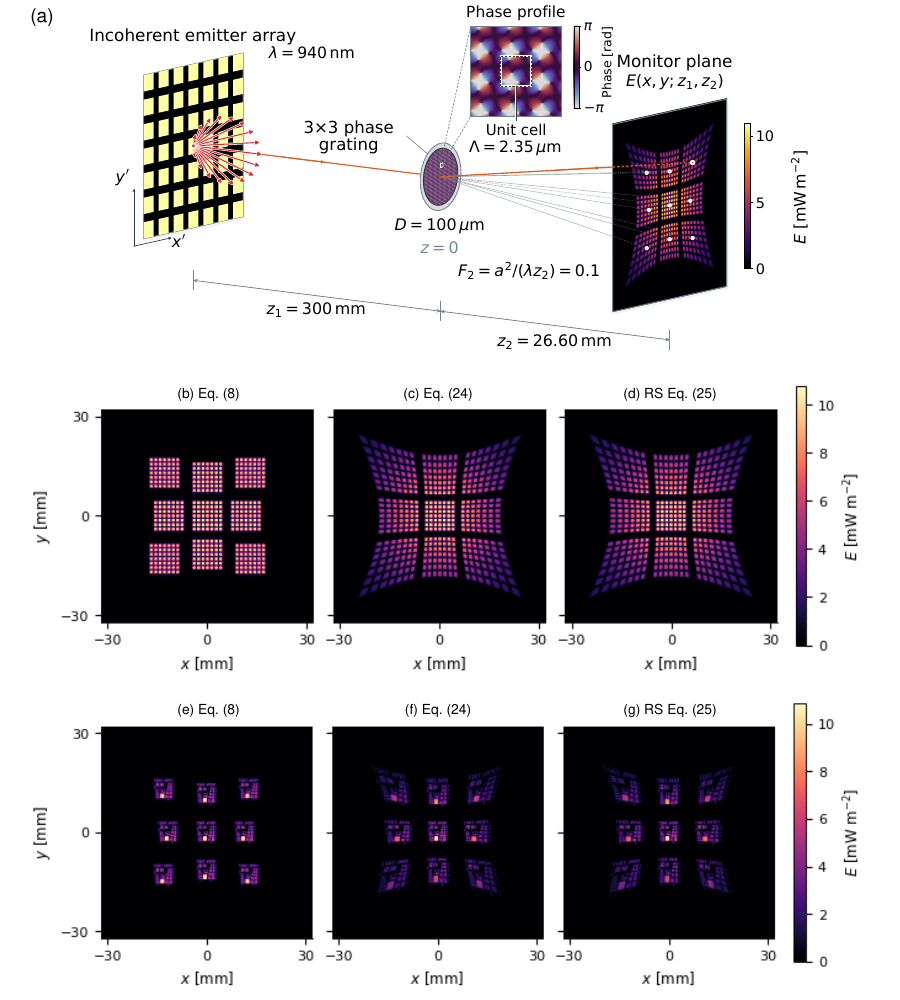}
\caption{Comparison of propagation models for finite-monitor imaging with a $3\times3$ phase grating.
(a) Emitter array, phase pupil, and monitor; the phase inset shows three periods along each axis and outlines one cell.
(b--d) Images of the $7\times7$ array of square emitters, computed with Eq.~\eqref{eq:psfconv}, Eq.~\eqref{eq:monitorconv}, and RS Eq.~\eqref{eq:monitordirect}, respectively.
(e--g) The same comparison for a USAF-shaped source.
The geometry is unchanged between rows; images within each row share a common absolute irradiance scale.
Phase and monitor maps in (a) are computed; distances are schematic.}
\label{fig:grating}
\end{figure}

We next use a $3\times3$ phase grating as the pupil and test how the models reproduce source images in its nine diffraction orders. Figure~\ref{fig:grating}(a) shows a $7\times7$ array of mutually incoherent square emitters at $z_1=300$~mm, each $10$~mm wide at $15$~mm pitch and radiance $10^4~\mathrm{W\,m^{-2}\,sr^{-1}}$. The circular pupil has a radius of $50~\mu$m and a grating period of $\Lambda=2.35~\mu$m at $\lambda=940$~nm. Its periodic cell places $93.95\%$ of Fourier-coefficient power in the nine desired orders, with $0.32\%$ peak-to-valley order nonuniformity. The $64\times64$~mm monitor lies at $z_2=26.60$~mm ($F_2=0.1$). We evaluate monitor power collection and image uniformity separately.

To explain the geometric distortion, we use the grating equation to find where each diffraction order reaches the monitor:
\begin{equation}
\alpha_{mn}=\frac{m\lambda}{\Lambda}-\alpha',\quad
\beta_{mn}=\frac{n\lambda}{\Lambda}-\beta',\qquad
(x_{mn},y_{mn})=
\frac{z_2(\alpha_{mn},\beta_{mn})}
{\sqrt{1-\alpha_{mn}^2-\beta_{mn}^2}}.
\label{eq:distortion}
\end{equation}
Here $m,n\in\{-1,0,1\}$, and $(\alpha',\beta')$ are defined in
Eq.~\eqref{eq:sourceangles}. The zero order is an inverted source
image, $(x_{00},y_{00})=(-Mx',-My')$. The other images stretch
and their grid lines curve [Fig.~\ref{fig:grating}(b--d)].
Diagonal orders span approximately $20^\circ$--$53^\circ$.
Equation~\eqref{eq:psfconv} shifts one response across the monitor,
so it misses these changes in image shape.

Equation~\eqref{eq:monitorconv} agrees with the RS square-source image to approximately $0.05\%$ in relative $L^2$ norm. RS irradiance decreases by about $79\%$ along a diagonal order's bright features. Dividing out only the final $c^4$ factor in Eq.~\eqref{eq:monitorconv} preserves the geometry but removes most of this decrease, identifying the importance of monitor radiometric weighting. Supplement~1, Section~\ref{supp:grating}, defines the cell and image metrics and gives registered profiles and numerical checks. The deformation and irradiance fall-off persist for other source emitter geometries, as illustrated by replacing the square-array with a USAF source over the same $100\times100$~mm plane [Fig.~\ref{fig:grating}(e--g)].

\subsection{Inverse design of a VCSEL-array diffuser}
\label{sec:vcsel}
\begin{figure}[!htbp]
\centering\includegraphics[width=\linewidth]{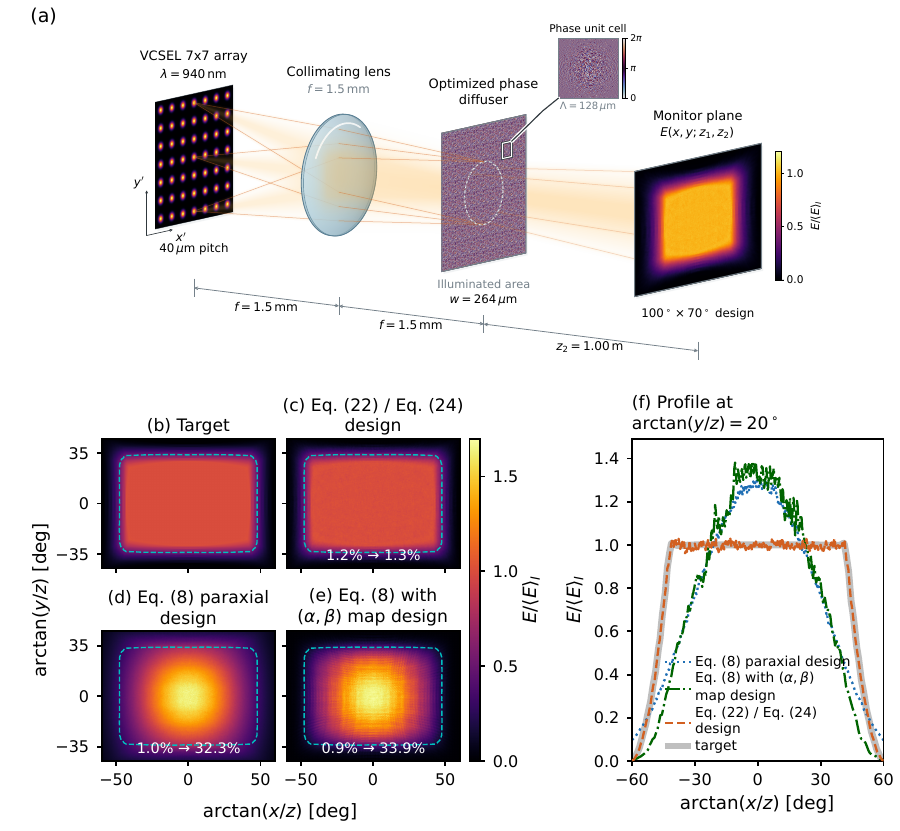}
\caption{VCSEL-array diffuser design at $z_2=1$~m. (a) Array, collimator, phase diffuser, and monitor; the inset shows one phase cell. (b) Target. (c) Nonparaxial design; the shared Eq.~\eqref{eq:common} / Eq.~\eqref{eq:monitorconv} label indicates their far-field agreement. (d,e) Paraxial and coordinate-corrected Eq.~\eqref{eq:psfconv} designs. (f) Profile at $\arctan(y/z_2)=20^\circ$. All maps and profiles are evaluated with Eq.~\eqref{eq:monitordirect} and normalized by each design's interior mean. Also, the maps share one color scale. Arrows compare predicted and RS-evaluated RMS nonuniformity for the first matched start. Cyan dashed contours mark the target's 50\% level.}
\label{fig:vcsel}
\end{figure}

This VCSEL phase optimization example illustrates the potential of the proposed models for inverse design. Specifically, we ask whether a phase diffuser designed with each forward model produces uniform wide-angle illumination. The setup consists of a $7\times7$ incoherent array at $940$~nm, with $40~\mu$m pitch, which is collimated onto a $128~\mu$m-period diffuser [Fig.~\ref{fig:vcsel}(a)]. The ideal sine-condition collimator has focal length $1.5$~mm, with the source and diffuser in its front and back focal planes. Each emitter contributes a Gaussian angular-radiance spot on a common illuminated area (Gaussian pupil) with $1/e^2$ intensity radius $w=264~\mu$m. We optimize the diffuser phase profile for uniform irradiance over a nominal $100^\circ\times70^\circ$ field at $z_2=1$~m, including finite-source edge roll-off; $F_2=w^2/(\lambda z_2)=0.074$.

The objective matches the target irradiance while penalizing exterior light and rewarding collection (Supplement~1, Section~\ref{supp:vcsel}). We compare four models: Eq.~\eqref{eq:psfconv}, a coordinate-corrected model that translates monitor responses without the monitor radiometric weight, and Eq.~\eqref{eq:monitorconv}; a separate run uses properly weighted Eq.~\eqref{eq:common}. All optimize the same scalar phase-screen parameterization from three matched starts for 800 Adam iterations \cite{KingmaBa2015}. Equation~\eqref{eq:monitorconv} uses the Gaussian order-lobe response described and independently checked in Section~\ref{supp:vcsel}.

We measure RMS nonuniformity as $U_{\rm RMS}=\operatorname{std}_I(E)/\langle E\rangle_I$ over the target interior $I$. After optimization, each model predicts approximately $0.9$--$1.2\%$. However, RS evaluation gives $32.29\%$ and $33.90\%$ for the paraxial and coordinate-corrected designs, in contrast with $1.27\%$ for both nonparaxial designs. The latter each collect $84.21\%$ of incident power in the nominal field. Although the coordinate-corrected design collects $92.15\%$, its bright center and dim corners miss the uniformity target [Fig.~\ref{fig:vcsel}(e)]. Therefore, accurate design requires both the coordinate mapping and radiometric weighting.

For both nonparaxial designs, Eqs.~\eqref{eq:common} and \eqref{eq:monitorconv} agree with independently checked RS images to about $0.05\%$ in relative $L^2$ norm. Thus, this example establishes a useful wide-angle design in which far-field and finite-distance predictions agree, showing why both nonparaxial geometry and the radiometric weighting in Eqs.~\eqref{eq:common} and \eqref{eq:monitorconv} are needed to achieve uniform illumination. The next example tests the benefit of finite-distance optimization using Eq.~\eqref{eq:monitorconv} over Eq.~\eqref{eq:common}.

\subsection{Finite-distance inverse design of a $3\times3$ flat-top array}
\label{sec:grating_inverse}
\begin{figure}[!htbp]
\centering\includegraphics[width=\linewidth]{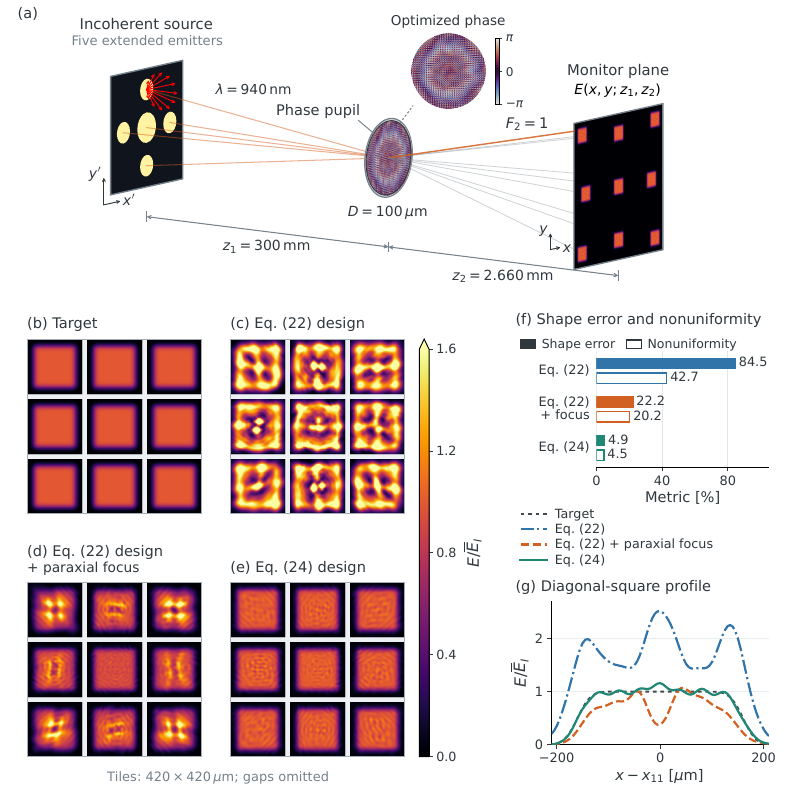}
\caption{Finite-distance design of a $3\times3$ flat-top array at $F_2=1$. (a) The optical setup consisting of five extended incoherent emitters, a circular phase pupil, and a monitor. (b) shows the target $T$ irradiance, while (c), (d) and (e) show the irradiance evaluated with full RS for the Eq. \eqref{eq:common} design, the Eq. \eqref{eq:common} design with paraxial focus added, and the Eq. \eqref{eq:monitorconv} design, respectively. Each tile spans $420\times420~\mu\mathrm m$; the gaps between tiles are omitted. (f) Target shape error $\epsilon_{\mathrm{shape}}=\|E/\overline E_I-T\|_2/\|T\|_2$ over all nine tiles and spatial nonuniformity $U$ over their interiors. (g) Horizontal profiles through the upper-right square. Paraxial refocusing adds the prescribed phase $-\pi(\xi^2+\eta^2)/(\lambda z_2)$.}
\label{fig:grating_inverse}
\end{figure}

In this example, we test finite-distance irradiance shaping. We keep the wavelength and pupil radius from Section~\ref{sec:grating}, set $z_2=2.660$~mm ($F_2=1$), and use five extended incoherent emitters at $z_1=300$~mm. These equal-radiance disks have central radius $0.8$~mm and outer radii $0.8/\sqrt2$~mm, with outer centers at $(\pm2,0)$ and $(0,\pm2)$~mm. We optimize a free-form phase across the pupil to produce a target irradiance profile consisting of nine $320~\mu$m-wide flat-top squares with smooth edges [Fig.~\ref{fig:grating_inverse}(b)]. The goal of this objective is to match flat-top square irradiance profiles while rewarding power collection and shape uniformity (Supplement~1, Section~\ref{supp:grating_inverse}). We compare designs based on Equations~\eqref{eq:common} and \eqref{eq:monitorconv}. A third proposed design comparator adds a nominal paraxial focusing phase $-\pi(\xi^2+\eta^2)/(\lambda z_2)$ to the optimized Eq.~\eqref{eq:common} design. Both models optimize the same phase pupil variables with the same objective and same number of Adam iterations.

After optimization, we evaluate the resulting designs using the full RS integral in Eq.~\eqref{eq:monitordirect}. We evaluate spatial nonuniformity as $U=\operatorname{std}_I(E)/\overline E_I$, where $I$ contains the interiors of all nine squares and $\overline E_I$ is their common mean. The Eq.~\eqref{eq:common} design gives $42.73\%$ nonuniformity. Adding paraxial focus to the Eq.~\eqref{eq:common} design reduces it to $20.24\%$. Optimization with Eq.~\eqref{eq:monitorconv} yields $4.50\%$ and reduces the target-shape error from $22.16\%$ to $4.90\%$ relative to refocusing, correcting distorted oblique spots. Thus, both measures in the Eq.~\eqref{eq:monitorconv} design improve by about $4.5\times$. 

For the Eq.~\eqref{eq:monitorconv} design, the predicted irradiance agrees with an independent, source-resolved RS reference to $0.27\%$ in the relative $L^2$ norm over the nine sampled tiles. Reference-refinement changes are far below the design gain, supporting the benefit of finite-distance optimization using  Eq.~\eqref{eq:monitorconv} over Eq.~\eqref{eq:common} for this example.

\FloatBarrier
\subsection{Computational speedup for prescribed illumination}
We compare Eq.~\eqref{eq:monitorconv} with source-by-source angular-spectrum (AS) propagation for a compact clear-pupil USAF benchmark ($F_1=0.02$, $F_2=0.10$). Both methods use identical source samples and weights. Each uses its fastest tested settings meeting the common $1\%$ relative $L^2$ irradiance-error requirement against the full RS reference.

Every call includes source construction, propagation setup, and detector mapping. Equation~\eqref{eq:monitorconv} rebuilds its coherent response, making these complete forward evaluations relevant to inverse design with a changing pupil and prescribed illumination.

\begin{figure}[!htbp]
\centering\includegraphics[width=\linewidth]{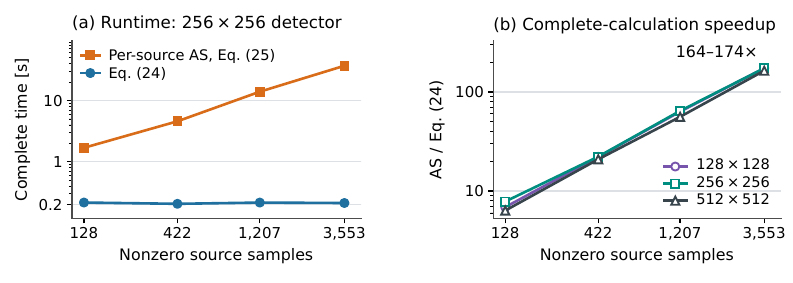}
\caption{Complete-calculation USAF benchmark. (a) Runtime for a $256\times256$ detector. (b) AS/Eq.~\eqref{eq:monitorconv} speedup for three detector grids. Times are medians of three calls, each rebuilding the coherent response. Supplement~1, Section~\ref{supp:timing}, gives the protocol.}
\label{fig:scaling}
\end{figure}

On the $256\times256$ detector, Eq.~\eqref{eq:monitorconv} remains near $0.21$~s as the source count increases from 128 to 3,553, while AS rises from $1.69$ to $36.93$~s [Fig.~\ref{fig:scaling}(a)]. At 3,553 samples, speedups are $164$--$174\times$ across the three detector grids [Fig.~\ref{fig:scaling}(b)].

\section{Discussion and conclusion}
We have developed a wide-angle convolution that combines absolute
radiometric normalization with a single reusable coherent response that builds on the derived GVS--source--PSF equivalence.
Direction-cosine mapping and the adopted scalar power prescription
then connect source covariance, Lambertian radiance, and monitor
irradiance on the same absolute scale. Angular convolution
is exact for the specified plane-wave mixture, including
separable angular transmission. At finite source and monitor
distances, comparisons with RS calculations assess the
incident-curvature and monitor-remapping approximations revealing substantial improvements in accuracy beyond the paraxial regime.

\begin{samepage}
The two inverse design examples illustrate complementary benefits.
The VCSEL diffuser shows why wide-angle design needs both
nonparaxial geometry and radiometric weighting, with the proposed Eqs.~\eqref{eq:common}
and \eqref{eq:monitorconv} giving consistent results near the far field.
For the finite-distance $3\times3$ flat-top array, Eq.~\eqref{eq:monitorconv} improves
spatial uniformity and irradiance shape over Eq.~\eqref{eq:common} designs both
with and without added paraxial focus. Reusing the coherent response substantially reduces computation time at matched accuracy. The resulting differentiable model
supports extended-source imaging and diffractive design,
and the accompanying code provides starting points for
reproducing and extending these examples for further applications.\par
\end{samepage}

\begin{backmatter}
\bmsection{Disclosures} The author declares no conflicts of interest.
\bmsection{Data availability}
The project repository on GitHub, \href{https://github.com/rafael-fuente/fast-incoherent-light-propagation}{https://github.com/rafael-fuente/fast-incoherent-light-propagation},
contains the numerical inputs, phase-grating data, executed validation notebooks,
and scripts used to generate the figures.
\bmsection{Supplemental document} Supplement~1 provides derivations, physical-limit tests, and numerical convergence procedures supporting the main results.
\end{backmatter}
\bibliography{references}

\ifdefined\OEtitle\else
  \def\htmlbibliography#1{\begin{thebibliography}{10}
\newcommand{\enquote}[1]{``#1''}

\bibitem{Schell1967}
A.~C. Schell, \enquote{{A technique for the determination of the radiation
  pattern of a partially coherent aperture},}
  \href{https://doi.org/10.1109/TAP.1967.1138864}{{\protect\JournalTitle{IEEE
  Transactions on Antennas and Propagation}}} \textbf{15}, 187--188 (1967).

\bibitem{Nugent1990}
K.~A. Nugent, \enquote{{A generalization of Schell's theorem},}
  \href{https://doi.org/10.1016/0030-4018(90)90066-3}{{\protect\JournalTitle{Optics
  Communications}}} \textbf{79}, 267--269 (1990).

\bibitem{Montoya2025}
M.~Montoya, M.~J. Lopera, Y.~Nie, \emph{et~al.}, \enquote{{Generalized Van
  Cittert-Zernike Schell propagator: an efficient algorithm for simulating
  partially coherent light},}
  \href{https://doi.org/10.1364/OE.571673}{{\protect\JournalTitle{Optics
  Express}}} \textbf{33}, 46313--46325 (2025).

\bibitem{Harvey1999}
J.~E. Harvey, C.~L. Vernold, A.~Krywonos, \emph{et~al.}, \enquote{{Diffracted
  radiance: a fundamental quantity in nonparaxial scalar diffraction theory},}
  \href{https://doi.org/10.1364/AO.38.006469}{{\protect\JournalTitle{Applied
  Optics}}} \textbf{38}, 6469--6481 (1999).

\bibitem{Harvey2000Errata}
J.~E. Harvey, C.~L. Vernold, A.~Krywonos, \emph{et~al.}, \enquote{{Diffracted
  radiance: a fundamental quantity in nonparaxial scalar diffraction theory:
  errata},}
  \href{https://doi.org/10.1364/AO.39.006374}{{\protect\JournalTitle{Applied
  Optics}}} \textbf{39}, 6374--6375 (2000).

\bibitem{Zissa2018}
D.~E. Zissa, \enquote{{Diffracted radiance not proven to be fundamental
  quantity in nonparaxial scalar diffraction theory: comment},}
  \href{https://doi.org/10.1364/AO.57.004997}{{\protect\JournalTitle{Applied
  Optics}}} \textbf{57}, 4997 (2018).

\bibitem{Smythe1947}
W.~R. Smythe, \enquote{{The Double Current Sheet in Diffraction},}
  \href{https://doi.org/10.1103/PhysRev.72.1066}{{\protect\JournalTitle{Physical
  Review}}} \textbf{72}, 1066--1070 (1947).

\bibitem{Blomstedt2016}
K.~Blomstedt, T.~Set{\"a}l{\"a}, J.~Tervo, \emph{et~al.},
  \enquote{{Vector-valued Lambertian fields and their sources},}
  \href{https://doi.org/10.1103/PhysRevA.93.053813}{{\protect\JournalTitle{Physical
  Review A}}} \textbf{93}, 053813 (2016).

\bibitem{Nicodemus1963}
F.~E. Nicodemus, \enquote{{Radiance},}
  \href{https://doi.org/10.1119/1.1969512}{{\protect\JournalTitle{American
  Journal of Physics}}} \textbf{31}, 368--377 (1963).

\bibitem{Walther1968}
A.~Walther, \enquote{{Radiometry and Coherence},}
  \href{https://doi.org/10.1364/JOSA.58.001256}{{\protect\JournalTitle{Journal
  of the Optical Society of America}}} \textbf{58}, 1256--1259 (1968).

\bibitem{Wolf1978}
E.~Wolf, \enquote{{Coherence and radiometry},}
  \href{https://doi.org/10.1364/JOSA.68.000006}{{\protect\JournalTitle{Journal
  of the Optical Society of America}}} \textbf{68}, 6--17 (1978).

\bibitem{CarterWolf1975}
W.~H. Carter and E.~Wolf, \enquote{{Coherence properties of lambertian and
  non-lambertian sources},}
  \href{https://doi.org/10.1364/JOSA.65.001067}{{\protect\JournalTitle{Journal
  of the Optical Society of America}}} \textbf{65}, 1067--1071 (1975).

\bibitem{CarterWolf1977}
W.~H. Carter and E.~Wolf, \enquote{{Coherence and radiometry with
  quasihomogeneous planar sources},}
  \href{https://doi.org/10.1364/JOSA.67.000785}{{\protect\JournalTitle{Journal
  of the Optical Society of America}}} \textbf{67}, 785--796 (1977).

\bibitem{ShenWang2006}
F.~Shen and A.~Wang, \enquote{{Fast-Fourier-transform based numerical
  integration method for the Rayleigh-Sommerfeld diffraction formula},}
  \href{https://doi.org/10.1364/AO.45.001102}{{\protect\JournalTitle{Applied
  Optics}}} \textbf{45}, 1102--1110 (2006).

\bibitem{VoelzRoggemann2009}
D.~G. Voelz and M.~C. Roggemann, \enquote{{Digital simulation of scalar optical
  diffraction: revisiting chirp function sampling criteria and consequences},}
  \href{https://doi.org/10.1364/AO.48.006132}{{\protect\JournalTitle{Applied
  Optics}}} \textbf{48}, 6132--6142 (2009).

\bibitem{MatsushimaShimobaba2009}
K.~Matsushima and T.~Shimobaba, \enquote{{Band-limited angular spectrum method
  for numerical simulation of free-space propagation in far and near fields},}
  \href{https://doi.org/10.1364/OE.17.019662}{{\protect\JournalTitle{Optics
  Express}}} \textbf{17}, 19662--19673 (2009).

\end{thebibliography}
}
  \let\bibliography\htmlbibliography
  \let\htmlcite\cite
  \let\htmlbibitem\bibitem
  \def\supphtmlbibitem#1{\htmlbibitem{supp-#1}}
  \let\bibitem\supphtmlbibitem
  \makeatletter
  \def\supphtmlcite#1{%
    \def\htmlkeys{}\def\htmlsep{}%
    \@for\htmlkey:=#1\do{%
      \edef\htmlkeys{\htmlkeys\htmlsep supp-\htmlkey}%
      \def\htmlsep{,}}%
    \expandafter\htmlcite\expandafter{\htmlkeys}}
  \makeatother
  \let\cite\supphtmlcite
\fi
\articletype{Supplemental Document}
\setcounter{section}{0}
\setcounter{equation}{0}
\setcounter{figure}{0}
\setcounter{table}{0}
\renewcommand{\thesection}{S\arabic{section}}
\renewcommand{\thesubsection}{\thesection.\arabic{subsection}}
\renewcommand{\theequation}{S\arabic{equation}}
\renewcommand{\thefigure}{S\arabic{figure}}
\renewcommand{\thetable}{S\arabic{table}}
\bibliographystyle{opticajnl-threeauthors}
\ifdefined\OEtitle
\title{Efficient convolution-based nonparaxial propagation of partially coherent light: supplementary material}
\author{Rafael de la Fuente Herrezuelo}
\email{rafael.fuente.herrezuelo@gmail.com}
\begin{abstract*}
\else
  \section*{Supplementary material}
\fi
This supplement explains the derivations and checks behind the main results. Sections \ref{supp:proof}--\ref{supp:angular} establish the normalized GVS--convolution equivalence, its radiometric meaning, and the approximations at finite source and monitor distances. Sections \ref{supp:checks}--\ref{supp:convergence} define the independent references and distinguish numerical error from propagation-model error. Sections \ref{supp:grating}--\ref{supp:timing} support the imaging, inverse-design, and timing examples. Detailed quadrature settings and run records accompany the notebooks.
\ifdefined\OEtitle\end{abstract*}\fi

\section{Normalized equivalence of GVS and the source--PSF convolution}
\label{supp:proof}
\subsection{Coherence and the pupil-pair reduction}
This derivation shows that GVS and the source--PSF convolution in main-text Section~\ref{sec:sourceconv} give the same irradiance, including its absolute normalization. We propagate the mutual intensity, then collect pupil pairs with the same separation. We use source coordinates $(x',y')$, pupil coordinates $\mathbf r=(\xi,\eta)$, and detector coordinates $(x,y)$. Distances $z_1$ and $z_2$ are positive, $M=z_2/z_1$, and $k=2\pi/\lambda$. The source covariance and Fourier convention are
\begin{align}
 J_0(\mathbf r'_1,\mathbf r'_2)&=\lambda^2L_0(\mathbf r'_1)\delta^{(2)}(\mathbf r'_1-\mathbf r'_2),\label{supp:eq:source}\\
 \widehat g(f_\xi,f_\eta)&=\iint g(\xi,\eta)e^{-i2\pi(f_\xi\xi+f_\eta\eta)}\,\mathrm d\xi\,\mathrm d\eta.\label{eq:fourier_convention}
\end{align}
Fresnel propagation gives $J_1\simeq e^{i\pi(|\mathbf r_1|^2-|\mathbf r_2|^2)/(\lambda z_1)}\Gamma(\mathbf r_1-\mathbf r_2)$, with
\begin{equation}
 \Gamma(\Delta\xi,\Delta\eta)=\frac{1}{z_1^2}\iint L_0(x',y')
 e^{-i2\pi(\Delta\xi x'+\Delta\eta y')/(\lambda z_1)}\,\mathrm dx'\,\mathrm dy'.
 \label{supp:eq:gamma}
\end{equation}
Combining the incident and outgoing Fresnel phases in
\begin{equation}
 Q(\xi,\eta)=\mathscr{P}(\xi,\eta)e^{i\pi(1/z_1+1/z_2)(\xi^2+\eta^2)/\lambda}
 \label{supp:eq:Qscreen}
\end{equation}
gives the pupil-pair integral
\begin{equation}
 E(\mathbf x)=\frac{1}{(\lambda z_2)^2}\iiiint Q(\mathbf r_1)Q^*(\mathbf r_2)
 \Gamma(\mathbf r_1-\mathbf r_2)e^{-i2\pi\mathbf x\cdot(\mathbf r_1-\mathbf r_2)/(\lambda z_2)}
 \,\mathrm d^2\mathbf r_1\,\mathrm d^2\mathbf r_2.\label{eq:pupil_pair}
\end{equation}
The irradiance thus contains a four-dimensional integral over pairs of pupil points. To reduce it, set $\mathbf r=(\mathbf r_1+\mathbf r_2)/2$ and $\Delta\mathbf r=\mathbf r_1-\mathbf r_2$, a change of variables with unit Jacobian. Integrating over the center coordinate gives
\begin{equation}
 C_Q(\Delta\mathbf r)=\iint Q(\mathbf r+\Delta\mathbf r/2)Q^*(\mathbf r-\Delta\mathbf r/2)\,\mathrm d^2\mathbf r
 =\mathcal F^{-1}\{|\widehat Q|^2\}(\Delta\mathbf r).\label{eq:pupil_autocorrelation}
\end{equation}
Only the separation remains to be integrated, giving the Fourier transform of $\Gamma C_Q$. This is the aperture-correlation reduction of Schell and Nugent \cite{Schell1967,Nugent1990}, and the structure of Montoya et al.'s Eq.~\href{https://doi.org/10.1364/OE.571673}{(10)} \cite{Montoya2025}:
\begin{equation}
E(x,y)=\frac{1}{(\lambda z_2)^2}
\mathcal F\!\left\{
\Gamma(\Delta\xi,\Delta\eta)\,
\mathcal F^{-1}\!\left\{|\widehat Q|^2\right\}
(\Delta\xi,\Delta\eta)
\right\}\!\left(\frac{x}{\lambda z_2},\frac{y}{\lambda z_2}\right).
 \label{eq:nested}
\end{equation}
We now rewrite this Fresnel expression as an ordinary source--PSF convolution.

\subsection{The source transform and area Jacobian}
The second Fourier transform makes the source coordinates explicit. Applying it to Eq.~\eqref{supp:eq:gamma} gives
\begin{align}
 \widehat\Gamma(f_x,f_y)
 &=\frac{1}{z_1^2}\iint L_0(x',y')\,
 \delta\!\left(f_x+\frac{x'}{\lambda z_1}\right)
 \delta\!\left(f_y+\frac{y'}{\lambda z_1}\right)\,\mathrm dx'\,\mathrm dy'\nonumber\\
 &=\lambda^2L_0(-\lambda z_1f_x,-\lambda z_1f_y).\label{eq:source_transform}
\end{align}
Each delta supplies a factor $\lambda z_1$. The minus signs account for the source inversion produced by two forward transforms. The product--convolution theorem converts Eq.~\eqref{eq:nested} to
\begin{align}
 E(x,y)=\frac{1}{(\lambda z_2)^2}\iint&\lambda^2L_0(-\lambda z_1\nu_x,-\lambda z_1\nu_y)\nonumber\\
 &\times\left|\widehat Q\!\left(\frac{x}{\lambda z_2}-\nu_x,\frac{y}{\lambda z_2}-\nu_y\right)\right|^2
 \,\mathrm d\nu_x\,\mathrm d\nu_y.\label{eq:frequency_convolution}
\end{align}
For a unit spatial delta at the original source plane, the coherent response defines $\operatorname{PSF}(x,y)=|h(x,y;0,0)|^2$. In the Fresnel model its closed form is
\begin{equation}
 \operatorname{PSF}(x,y)=\frac{1}{\lambda^4z_1^2z_2^2}
 \left|\mathcal F\{Q\}\!\left(\frac{x}{\lambda z_2},\frac{y}{\lambda z_2}\right)\right|^2.
 \label{eq:fresnelpsf}
\end{equation}
Substitute $x'=-\lambda z_1\nu_x$, $y'=-\lambda z_1\nu_y$ to obtain
\begin{equation}
 E(x,y)=\iint\lambda^2L_0(x',y')\operatorname{PSF}(x+Mx',y+My')\,\mathrm dx'\,\mathrm dy'.
 \label{eq:source_sum}
\end{equation}
Finally, $u=-Mx'$, $v=-My'$ gives $\mathrm dx'\,\mathrm dy'=M^{-2}\mathrm du\,\mathrm dv$, hence
\begin{equation}
 E(x,y)=\frac{1}{M^2}\left[\lambda^2L_0\!\left(-\frac{x}{M},-\frac{y}{M}\right)\otimes\operatorname{PSF}(x,y)\right].
 \label{eq:source_conv}
\end{equation}
Here $(g\otimes h)(x,y)=\iint g(u,v)h(x-u,y-v)\,\mathrm du\,\mathrm dv$. The source is inverted and magnified by $M$, while the factor $M^{-2}$ accounts for its mapped area. These coordinate changes supply the absolute normalization of the GVS--source--PSF equivalence.

\subsection{Coherent-engine normalization and scope}
To use a numerical coherent propagator in the convolution, its response must have the normalization of Eq.~\eqref{eq:fresnelpsf}. Start with unit pupil-plane amplitude and the incident phase $\exp[i\pi(\xi^2+\eta^2)/(\lambda z_1)]$, multiply by the pupil, and propagate to the monitor. If the resulting intensity is $E_{\mathrm{sys}}$, then $\operatorname{PSF}=E_{\mathrm{sys}}/(\lambda z_1)^2$. Substitution into Eq.~\eqref{eq:source_conv} gives $E=z_2^{-2}[L_0(-x/M,-y/M)\otimes E_{\mathrm{sys}}]$, the normalization used in the implementation.

A reusable PSF requires each emitter's intensity response to be a translated copy of the central response, as in Eq.~\eqref{eq:source_sum}. An angular-spectrum engine can calculate that central response without a Fresnel propagation approximation, but the translation rule remains a separate assumption. The main article therefore derives the angular shift rule for incident plane waves and introduces common incident curvature when the source is at finite distance.

For the Fresnel closed form, Parseval's identity gives
\begin{align}
 \iint\operatorname{PSF}(x,y)\,\mathrm dx\,\mathrm dy
 &=\frac{1}{\lambda^2z_1^2}\iint|\mathscr{P}(\xi,\eta)|^2\,\mathrm d\xi\,\mathrm d\eta,\label{eq:psf_integral}\\
 \iint E(x,y)\,\mathrm dx\,\mathrm dy
 &=\Gamma(0,0)\iint|\mathscr{P}(\xi,\eta)|^2\,\mathrm d\xi\,\mathrm d\eta.\label{eq:fresnel_power}
\end{align}
These integrals cover the full paraxial output plane; a finite crop collects only part of the power.

\paragraph{Numerical sampling and checks.}
We write Eq.~\eqref{eq:source_conv} as
$E=G\otimes\operatorname{PSF}$, where
$G(x,y)=\lambda^2M^{-2}L_0(-x/M,-y/M)$.
Both arrays are sampled on the same coordinate grid, and the discrete
convolution includes the cell area $h_xh_y$. For retained array lengths
$N_G$ and $N_P$, the FFT size must satisfy
\begin{equation*}
N_{\mathrm{FFT}}\geq N_G+N_P-1
\end{equation*}
along each axis. Zero-padding to this size gives the full linear
convolution without circular wraparound. Refining the pupil grid and
enlarging the retained PSF window provide checks on the sampling and
truncation of the coherent response.

For integrable $G$ and $\operatorname{PSF}$, integration over the full
plane gives
\begin{equation*}
\iint_{\mathbb R^2}E\,\mathrm dx\,\mathrm dy
=
\left(\iint_{\mathbb R^2}G\,\mathrm dx\,\mathrm dy\right)
\left(\iint_{\mathbb R^2}\operatorname{PSF}\,\mathrm dx\,\mathrm dy\right).
\end{equation*}
This identity provides a simple check of the normalization and
integration weights. If the output is cropped, or an angular
convolution is restricted to the physical direction-cosine disk,
the excluded power must be accounted for separately.

Let $\epsilon_P$ denote the fraction of the full-plane PSF integral
outside the retained window. When both functions are nonnegative and
the source map is unchanged, $\epsilon_P$ also gives the fraction of
full-plane output power lost by truncating the PSF. Pointwise accuracy,
grid convergence, and the accuracy of the optical model require
separate checks.

For inverse design, derivatives must include every step that depends
on the design parameters. If a parameter $p$ changes the source map
$G$, the derivative includes
$(\partial G/\partial p)\otimes\operatorname{PSF}$
in addition to the pupil-response term given in the main article.
Resampling and cropping must also be differentiated when they depend
on the design parameters.

\section{Angular normalization, physical limits, and curvature}
\label{supp:angular}
\begingroup
\setlength{\abovedisplayskip}{6pt plus 2pt minus 2pt}
\setlength{\belowdisplayskip}{4pt plus 2pt minus 2pt}
\setlength{\abovedisplayshortskip}{0pt plus 2pt}
\setlength{\belowdisplayshortskip}{3pt plus 2pt minus 2pt}
\setlength{\textfloatsep}{12pt plus 2pt minus 2pt}
\setlength{\intextsep}{10pt plus 2pt minus 2pt}
\captionsetup{skip=6pt}
\predisplaypenalty=1000
\subsection{Outgoing Rayleigh--Sommerfeld field and plane flux}
\label{supp:angular_power}

The angular convolution in main-text Section~\ref{sec:angular} requires a consistent relation between the scalar field and radiated power. We derive the far-field response and state the normalization that supplies this relation. Let $U_b(\xi,\eta)$ be the
field immediately after the pupil. Write the observation point as
$R\widehat{\mathbf s}$, where $\widehat{\mathbf s}=(\alpha,\beta,c)$
and $c=\sqrt{1-\alpha^2-\beta^2}>0$. In the far field, the
Rayleigh--Sommerfeld kernel gives
\begin{equation}\label{eq:RSfar}
U(R\widehat{\mathbf s})
\simeq
\frac{ce^{ikR}}{i\lambda R}
\iint U_b(\xi,\eta)e^{-ik(\alpha\xi+\beta\eta)}
\,\mathrm d\xi\,\mathrm d\eta.
\end{equation}
For a pupil contained within a circle of radius $a$, this approximation
requires $ka^2/R\ll1$, $a/R\ll1$, and $kR\gg1$. No small-angle
approximation is made for the observation direction. Using the pupil
mutual intensity
$J_b(\mathbf r_1,\mathbf r_2)
=\langle U_b(\mathbf r_1)U_b^*(\mathbf r_2)\rangle$,
define
\begin{equation}\label{eq:B}
\mathcal B(\alpha,\beta)
=
\iiiint J_b(\mathbf r_1,\mathbf r_2)
e^{-ik[\alpha(\xi_1-\xi_2)+\beta(\eta_1-\eta_2)]}
\,\mathrm d^2\mathbf r_1\,\mathrm d^2\mathbf r_2.
\end{equation}
Equation~\eqref{eq:RSfar} then gives
$\langle|U|^2\rangle\simeq c^2\mathcal B/(\lambda^2R^2)$.

The time-averaged Poynting vector is
$\mathbf{S}=\frac12\operatorname{Re}(\mathbf E\times\mathbf H^*)$.
With brackets denoting the ensemble average, the radiant intensity is
$\mathrm dP/\mathrm d\Omega
=\lim_{R\to\infty}R^2\langle\mathbf{S}\rangle
\cdot\widehat{\mathbf s}$.
For radial far-field flux, projection onto the pupil normal gives
\begin{equation*}
\langle\mathbf{S}\rangle\cdot\widehat{\mathbf z}
=
c\langle\mathbf{S}\rangle\cdot\widehat{\mathbf s}.
\end{equation*}
We adopt the plane-flux normalization underlying Harvey et al.'s
angular power relation \cite{Harvey1999,Harvey2000Errata}:
\begin{equation}\label{eq:identification}
\langle|U|^2\rangle
\equiv
\langle\mathbf{S}\rangle\cdot\widehat{\mathbf z}.
\end{equation}
With this convention, $|U|^2$ represents flux through a plane parallel
to the pupil. For a locally transverse outgoing electromagnetic field
in a lossless medium, this corresponds to
$|U|^2=Kc|\mathbf E|^2$, where
$K=\frac12\sqrt{\varepsilon/\mu}$ is absorbed into the scalar
normalization. Combining Eqs.~\eqref{eq:RSfar}
and~\eqref{eq:identification} gives
\begin{equation}\label{eq:angularpower}
\frac{\mathrm dP}{\mathrm d\Omega}(\alpha,\beta)
=
\lim_{R\to\infty}\frac{R^2}{c}
\langle|U(R\widehat{\mathbf s})|^2\rangle
=
\frac{c}{\lambda^2}\mathcal B(\alpha,\beta).
\end{equation}
This recovers Eq.~\eqref{eq:angularoutput} of the main article in Harvey et al.'s scalar
diffracted-radiance convention \cite{Harvey1999,Harvey2000Errata}.
The field correlation contains a factor $c^2$; converting plane flux
to radiant intensity divides by $c$, leaving a single cosine.
The scalar quantity $S=\mathcal B/\lambda^2$ therefore satisfies
$\mathrm dP/\mathrm d\Omega=cS$. Integrating $S$ over
$\mathrm d\alpha\,\mathrm d\beta=c\,\mathrm d\Omega$ gives the
radiant power.

Equation~\eqref{eq:identification} is the radiometric assumption used
throughout this work. Harvey et al.'s erratum revises the original
equal-integrals argument, and Zissa questions the hemisphere-flux
interpretation based on $\langle|U|^2\rangle/c$
\cite{Harvey2000Errata,Zissa2018}. Our normalization checks establish
consistency with the adopted convention, while the RS comparisons
test propagation accuracy within it. These checks do not provide an
independent electromagnetic validation.
The vector-aperture benchmark below provides a physical motivation
for this choice: for normally incident unpolarized illumination
within the prescribed tangential-field model, the single cosine
is closer to the vector angular factor than the squared cosine
for every $0<c<1$.
Comparisons between scalar conventions must account for this
normalization choice. For a fixed pupil-plane irradiance $E_0$ and
incident cosine $c_0$, the ray-normal convention uses an incident
amplitude $\sqrt{E_0/c_0}$ and retains the outgoing factor $c^2$.
Its radiant intensity is therefore related to the plane-flux result
by $R_{\mathrm{ray}}=(c/c_0)R_{\mathrm{plane}}$.
For an extended source, the incident factor $1/c'$ must be included
for each direction inside the source integral.

\paragraph{Covariance coefficient, radiance, and exitance.}
We can now determine the physical meaning of a delta covariance. Denote its nonnegative coefficient by $C_0$:
\begin{equation*}
J_0(\mathbf r'_1,\mathbf r'_2)
=C_0(\mathbf r'_1)\delta^{(2)}(\mathbf r'_1-\mathbf r'_2),\qquad C_0\geq0.
\end{equation*}
Substituting this covariance directly into Eq.~\eqref{eq:angularpower} removes the Fourier phase and gives
\begin{equation*}
\frac{\mathrm d^2P_0}{\mathrm dA'\,\mathrm d\Omega}
=\frac{c}{\lambda^2}C_0(\mathbf r'),\qquad
L(\mathbf r';\theta,\varphi)
=\frac1c\frac{\mathrm d^2P_0}{\mathrm dA'\,\mathrm d\Omega}
=\frac{C_0(\mathbf r')}{\lambda^2}.
\end{equation*}
The radiance is therefore angle-independent. Integrating over the forward hemisphere gives
\begin{equation*}
I_0(\mathbf r')=\int_{\Omega_+}L c\,\mathrm d\Omega
=\frac{\pi C_0(\mathbf r')}{\lambda^2},\qquad
C_0=\lambda^2L_0=\frac{\lambda^2}{\pi}I_0.
\end{equation*}
Thus $L_0=C_0/\lambda^2$ is the angle-independent radiance and $I_0=\pi L_0$ is the forward hemispherical exitance. The same calculation fixes both the Lambertian angular law and its absolute power scale.

\begin{figure}[!t]
\centering
\includegraphics[width=\linewidth]{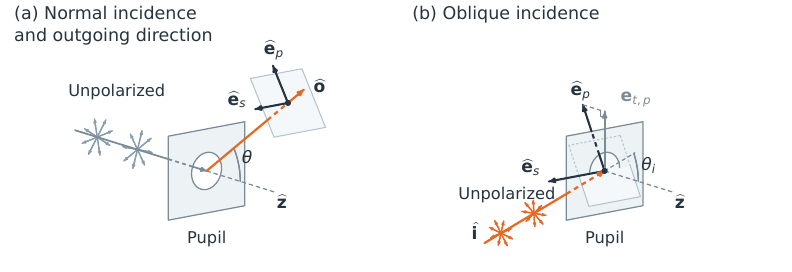}
\captionsetup{width=\linewidth}
\caption{Polarization geometry for (a) normal and (b) oblique incidence.
Black arrows are unit vectors; gray $\mathbf e_{t,p}$ is the tangential projection. Spokes indicate unpolarized light.}
\label{fig:supp_polarization}
\end{figure}

\paragraph{Vector-aperture motivation for unpolarized illumination.}
A vector-aperture comparison helps explain the single-cosine choice
[Fig.~\ref{fig:supp_polarization}(a)]. Using Smythe's representation for an aperture in a perfectly conducting plane screen, we prescribe the tangential electric field through the same scalar pupil for both incident polarizations \cite{Smythe1947}.
With $\widehat{\mathbf o}=(\sin\theta,0,c)$, $c=\cos\theta$, $\widehat{\mathbf e}_s=\widehat{\mathbf y}$, and $\widehat{\mathbf e}_p=(c,0,-\sin\theta)$, the far-field projection of the Fourier transform $\widetilde{\mathbf E}_t=\widetilde E_{t,p}\widehat{\mathbf x}+\widetilde E_{t,s}\widehat{\mathbf y}$ is, apart from common propagation constants,
\begin{align*}
\mathbf E_\infty\propto
\widehat{\mathbf o}\times(\widehat{\mathbf z}\times\widetilde{\mathbf E}_t)
&=-\widetilde E_{t,p}\widehat{\mathbf e}_p-c\widetilde E_{t,s}\widehat{\mathbf e}_s,\\
\frac{\mathrm dP}{\mathrm d\Omega}&\propto
\langle|\widetilde E_{t,p}|^2\rangle+c^2\langle|\widetilde E_{t,s}|^2\rangle.
\end{align*}
Under normal incidence, unpolarized illumination and a common scalar pupil give equal, uncorrelated tangential spectra:
\begin{equation*}
\langle|\widetilde E_{t,s}|^2\rangle
=\langle|\widetilde E_{t,p}|^2\rangle=\frac{\Phi}{2},
\qquad
\langle\widetilde E_{t,s}\widetilde E_{t,p}^*\rangle=0.
\end{equation*}
Averaging the two polarizations gives the factor $f_{\mathrm{vec}}=(1+c^2)/2$. The single-cosine factor $f_{\mathrm H}=c$ is the geometric mean of the two polarization factors, $c=\sqrt{c^2\cdot1}$; the squared-cosine factor $f_{\mathrm{sq}}=c^2$ equals the $s$-polarized response in this tangential-electric-field representation. Their differences from the vector average are
\begin{align*}
f_{\mathrm{vec}}-f_{\mathrm H}
&=\frac{(1-c)^2}{2}=\frac{\theta^4}{8}+O(\theta^6),\\
f_{\mathrm{vec}}-f_{\mathrm{sq}}
&=\frac{1-c^2}{2}=\frac{\theta^2}{2}+O(\theta^4),
\end{align*}
where angles in the expansions are in radians. The following values report relative underestimates, $100(f_{\mathrm{vec}}-f)/f_{\mathrm{vec}}$, for the normal-incidence benchmark:
\begin{center}
\begin{tabular}{rrrr}
\toprule
$\theta$ & $f_{\mathrm{vec}}$ & Single cosine [\%] & Squared cosine [\%]\\
\midrule
$20^\circ$ & 0.9415 & 0.2 & 6.2\\
$40^\circ$ & 0.7934 & 3.4 & 26\\
$60^\circ$ & 0.6250 & 20 & 60\\
\bottomrule
\end{tabular}
\end{center}

For a delta-correlated tangential electric field with equal, uncorrelated components, the same benchmark gives $L(\theta)/L(0)=(1+\cos^2\theta)/(2\cos\theta)=1+\theta^4/8+O(\theta^6)$. The electromagnetic Lambertian sources with wavelength-scale correlations are discussed by Blomstedt et al.~\cite{Blomstedt2016}.

\paragraph{Equal polarization powers at oblique incidence.}
For an extended unpolarized source, equal and uncorrelated physical $s$ and $p$ powers are imposed separately in every incident direction; the angular radiance itself need not be uniform.
In Fig.~\ref{fig:supp_polarization}(b), $\widehat{\mathbf e}_s$ lies in
the pupil plane, whereas $\widehat{\mathbf e}_p$ projects to
$\mathbf e_{t,p}$ of length $\cos\theta_i$, giving unequal tangential variances.
With $c_j=\cos\theta_j$ and $\mathbf q_j=\sin\theta_j(\cos\varphi_j,\sin\varphi_j)$ for actual propagation directions $j=i,o$, impedance and incident plane-irradiance normalization give the tangential covariance and outgoing projection metric
\begin{equation*}
\mathbf C_{t,i}=\frac{\mathbf I_2-\mathbf q_i\mathbf q_i^{\mathrm T}}{2c_i},\qquad \mathbf M_o=c_o^2\mathbf I_2+\mathbf q_o\mathbf q_o^{\mathrm T}.
\end{equation*}
After removing the common pupil Fourier-intensity factor, the trace $\operatorname{tr}(\mathbf M_o\mathbf C_{t,i})$ gives
\begin{equation*}
f_{\mathrm{vec}}(i,o)=c_o+\frac{(c_o-c_i)^2+
\sin^2\theta_i\sin^2\theta_o\sin^2(\varphi_o-\varphi_i)}{2c_i}.
\end{equation*}
Both \(f_{\rm vec}\) and \(f_{\rm H}=c_o\) are normalized to the same incident irradiance on the pupil plane. For coplanar incident and outgoing directions, the two polarization power factors are \(c_o^2/c_i\) and \(c_i\). The vector result is their arithmetic mean, while the scalar factor \(c_o\) is their geometric mean. The same relationship holds at arbitrary relative azimuth for the two orthogonal incident polarizations that give the smallest and largest power factors. The product of these factors is \(\det(\mathbf M_o)\det(2\mathbf C_{t,i})=c_o^2\).

The scalar and vector factors agree when the propagation direction is unchanged. For small departures from that direction, their difference is second order at oblique incidence and fourth order at normal incidence.

For an extended incoherent source, contributions from different incident directions add with nonnegative weights. If the relative error against the vector benchmark is bounded by the same value for every contributing direction, that bound also applies to the summed angular power. These results motivate a single-cosine approximation for unpolarized illumination within the prescribed tangential-field model.

\subsection{Incident propagation, the Jacobian, and irradiance}
\label{supp:incident}
Next we map the source into the angular coordinates used by the convolution. For a source point $(x',y',-z_1)$, write $R'=\sqrt{z_1^2+x'^2+y'^2}$, $\alpha'=x'/R'$, $\beta'=y'/R'$, and $c'=z_1/R'$. The primed direction points toward the emitter. The inverse map and its derivative are
\begin{equation}\label{eq:sourcemap}
(x',y')=\frac{z_1}{c'}(\alpha',\beta'),\qquad L_{\mathrm{in}}(\alpha',\beta')=L_0\!\left(\frac{z_1\alpha'}{c'},\frac{z_1\beta'}{c'}\right),
\end{equation}
\begin{equation}\label{eq:jacobian}
\frac{\partial(x',y')}{\partial(\alpha',\beta')}=\frac{z_1}{c'^3}
\begin{pmatrix}1-\beta'^2&\alpha'\beta'\\\alpha'\beta'&1-\alpha'^2\end{pmatrix},\qquad
\left|\det\frac{\partial(x',y')}{\partial(\alpha',\beta')}\right|=\frac{z_1^2}{c'^4}.
\end{equation}
Evaluate the incident kernel amplitude at the pupil center and retain its linear phase. The covariance coefficient $\lambda^2L_0$ cancels the kernel's $\lambda^{-2}$, giving
\begin{align}\label{supp:eq:angularcoherence}
\Gamma(\Delta\xi,\Delta\eta)&\simeq\iint L_0(x',y')\frac{z_1^2}{R'^4}e^{-ik(\alpha'\Delta\xi+\beta'\Delta\eta)}\,\mathrm dx'\,\mathrm dy'\nonumber\\
&=\iint_{\mathcal D} L_{\mathrm{in}}(\alpha',\beta')e^{-ik(\alpha'\Delta\xi+\beta'\Delta\eta)}\,\mathrm d\alpha'\,\mathrm d\beta'.
\end{align}
Here $\mathcal D$ is the unit disk. The source-area Jacobian cancels the central propagation weight: $(z_1^2/R'^4)(z_1^2/c'^4)=1$. The first line neglects amplitude variation over the pupil and the relative $1/(kR')^2$ contribution to the central squared amplitude. The second line also defines an incoherent plane-wave mixture directly. For a finite source, the common-curvature approximation supplies its incident wavefront.

The cancellation has a simple radiometric interpretation. From Eq.~\eqref{eq:angularpower}, the delta-correlated source emits $\mathrm dP_0/\mathrm d\Omega=c'\iint L_0\,\mathrm dA'$, as derived in the main article. A source element therefore illuminates a parallel receiving plane with
\begin{equation}\label{eq:incidentpower}
\mathrm dE=L_0\frac{\cos\theta_{\mathrm{source}}\cos\theta_{\mathrm{receiver}}}{R'^2}\,\mathrm dA'=L_0\frac{z_1^2}{R'^4}\,\mathrm dA'.
\end{equation}
The two cosines are equal because the planes are parallel. Thus the incoming kernel weight is precisely the geometric plane irradiance of a Lambertian element. Using $\mathrm d\alpha'\,\mathrm d\beta'=c'\,\mathrm d\Omega'$ gives
\begin{equation}\label{eq:irradiance}
E=\Gamma(0,0)=\iint_{\mathcal D}L_{\mathrm{in}}\,\mathrm d\alpha'\,\mathrm d\beta'=\iint_{\Omega_+}L_{\mathrm{in}}c'\,\mathrm d\Omega'.
\end{equation}
The angular integral therefore returns the incident pupil-plane irradiance directly.

\paragraph{Radiometric input from nonplanar sources.}
The same construction can be used for a curved or irregular emitting surface \cite{Nicodemus1963,Walther1968,Wolf1978}. At a fixed pupil point $\mathbf r$, let a source element $\mathrm dA_s$ lie at distance $R_s$, with source and receiver angles $\theta_s$ and $\theta_r$ (main-text Fig.~\ref{fig:nonplanar_source}(a)). In a homogeneous, lossless medium, the incident radiance equals the emitted radiance along the connecting ray [Fig.~\ref{fig:supp_radiance_curvature}(a)]. Hence
\begin{align*}
\mathrm dE&=L_s\frac{\cos\theta_s\cos\theta_r}{R_s^2}\,\mathrm dA_s
=L_{\mathrm{in}}\cos\theta_r\,\mathrm d\Omega'
=L_{\mathrm{in}}\,\mathrm d\alpha'\,\mathrm d\beta',\\
\mathrm d\Omega'&=\frac{\cos\theta_s}{R_s^2}\,\mathrm dA_s.
\end{align*}
Trace each incident direction back to the visible source surface and assign its emitted radiance to that ray. This gives $L_{\mathrm{in}}(\mathbf r,\alpha',\beta')$ at each pupil point. A single convolution uses the value at the pupil center when its spatial variation is negligible. At finite distance, we must also represent the incident wavefront curvature. For a source point at direction $\mathbf q'=(\alpha',\beta')$ and distance $R'(\mathbf q')$ from that center [Fig.~\ref{fig:supp_radiance_curvature}(b)], the quadratic incident path term is
\begin{equation*}
\Delta R^{(2)}(\mathbf r;\mathbf q')
=\frac{|\mathbf r|^2-(\mathbf q'\cdot\mathbf r)^2}{2R'(\mathbf q')}.
\end{equation*}
Its numerator is the squared displacement perpendicular to the central source direction. The equivalent plane in main-text Fig.~\ref{fig:nonplanar_source}(b) preserves directional radiance at the pupil center. Reusing one coherent response additionally requires the curvature above to vary little across the source directions, just as for a finite planar source.

\begin{figure}[!htbp]
\centering
\includegraphics[width=\linewidth]{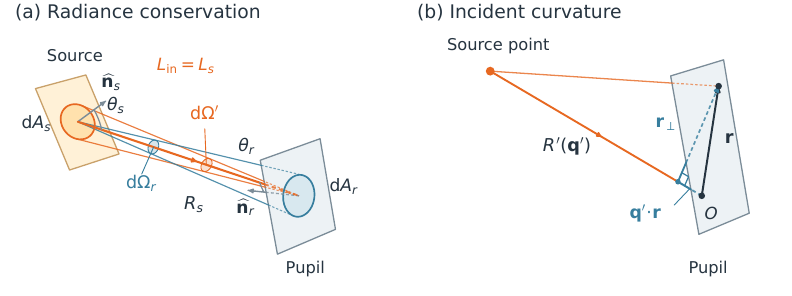}
\caption{Radiance transport and incident-wavefront geometry.
(a) Source and pupil elements subtend reciprocal solid angles
$\mathrm d\Omega'$ and $\mathrm d\Omega_r$, respectively.
Radiance is unchanged along their connecting ray in a homogeneous,
lossless medium.
(b) A source point lies at distance $R'(\mathbf q')$ from pupil center $O$.
Resolving the pupil displacement $\mathbf r$ along the central source
direction gives parallel length $\mathbf q'\cdot\mathbf r$.}
\label{fig:supp_radiance_curvature}
\end{figure}

\subsection{Angular nested transform and convolution}
With the angular illumination now specified, the same pupil-pair reduction used in Section~\ref{supp:proof} gives the angular convolution. Let $Q$ be the effective pupil: $Q=\mathscr{P}$ for illumination from infinity and $Q=\mathscr{P}\exp[i\pi(\xi^2+\eta^2)/(\lambda z_1)]$ in the common-curvature model. Writing $J_b(\mathbf r_1,\mathbf r_2)=Q(\mathbf r_1)Q^*(\mathbf r_2)\Gamma(\mathbf r_1-\mathbf r_2)$ in Eq.~\eqref{eq:angularpower} and integrating over the mean pupil coordinate yields
\begin{equation}\label{eq:angularnested}
S(\alpha,\beta)=\frac1{\lambda^2}\mathcal F\!\left\{\Gamma\mathcal F^{-1}[|\widehat Q|^2]\right\}\!\left(\frac\alpha\lambda,\frac\beta\lambda\right).
\end{equation}
This is the angular counterpart of Eq.~\eqref{eq:nested}. Fourier transforming Eq.~\eqref{supp:eq:angularcoherence} gives $\widehat\Gamma(f_\xi,f_\eta)=\lambda^2L_{\mathrm{in}}(-\lambda f_\xi,-\lambda f_\eta)$, with $L_{\mathrm{in}}$ extended by zero outside $\mathcal D$. The product--convolution theorem and the frequency change $(u,v)=\lambda(f_\xi,f_\eta)$ then give
\begin{equation}\label{eq:angularpsf}
H_Q(\alpha,\beta)=\frac1{\lambda^2}\left|\widehat Q\!\left(\frac\alpha\lambda,\frac\beta\lambda\right)\right|^2,
\end{equation}
\begin{align}\label{supp:eq:angularconv}
S(\alpha,\beta)&=\iint_{\mathcal D}L_{\mathrm{in}}(\alpha',\beta')H_Q(\alpha+\alpha',\beta+\beta')\,\mathrm d\alpha'\,\mathrm d\beta'\nonumber\\
&=L_{\mathrm{in}}(-\alpha,-\beta)\otimes H_Q(\alpha,\beta).
\end{align}
The wavelength factors cancel the two-dimensional frequency Jacobian. The nested transform and convolution are therefore two evaluations of the same model. Section~\ref{supp:convergence} tests that equivalence at infinity and separates it from pupil discretization. The coherent Fourier response must be available at the sums of incident and outgoing direction cosines. Zero padding implements linear convolution; enlarging the angular window checks the retained support. Select the outgoing propagating disk after convolution, keeping the original normalization.

\subsection{Plane-wave and transparent-plane limits}
A single plane wave and a transparent pupil test whether the angular convolution recovers coherent diffraction and preserves radiance. A plane wave occupies one incident direction, $(\alpha',\beta')=(\alpha_0,\beta_0)$, pointing toward the source. It propagates along $(-\alpha_0,-\beta_0,c_0)$, where $c_0=\sqrt{1-\alpha_0^2-\beta_0^2}$. Figure~\ref{fig:supp_transparent_plane} shows both descriptions of the same ray. With irradiance $E_0$ on the pupil plane, the incident radiance and mutual coherence are
\begin{align}
L_{\mathrm{in}}(\alpha',\beta')&=E_0\delta(\alpha'-\alpha_0)\delta(\beta'-\beta_0)=\frac{E_0}{c_0}\delta_\Omega(\Omega'-\Omega'_0),\label{eq:planewavesource}\\
\Gamma(\Delta\xi,\Delta\eta)&=E_0e^{-ik(\alpha_0\Delta\xi+\beta_0\Delta\eta)}.\label{eq:planewavecoherence}
\end{align}
Here $\Omega'_0$ is the selected direction toward the source, and $\delta_\Omega$ has unit integral over solid angle. Thus $E_0=\iint_{\mathcal D}L_{\mathrm{in}}\,\mathrm d\alpha'\,\mathrm d\beta'=\iint_{\Omega_+}L_{\mathrm{in}}c'\,\mathrm d\Omega'$, while the irradiance normal to the ray is $E_0/c_0$. Substituting Eq.~\eqref{eq:planewavesource} into Eq.~\eqref{supp:eq:angularconv} recovers the coherent diffraction pattern
\begin{equation}\label{eq:planewave}
\frac{\mathrm dP}{\mathrm d\Omega}(\alpha,\beta)=\frac{E_0c}{\lambda^2}\left|\widehat{\mathscr{P}}\!\left(\frac{\alpha+\alpha_0}{\lambda},\frac{\beta+\beta_0}{\lambda}\right)\right|^2.
\end{equation}
For comparison, a localized planar emitter of forward power $P_s$ has $L_0(\mathbf r')=(P_s/\pi)\delta^{(2)}(\mathbf r'-\mathbf r'_s)$ and gives pupil-center irradiance $E_0=(P_s/\pi)z_1^2/R_s'^4$.

A fully transparent plane ($\mathscr{P}=1$) transmits the wave unchanged. To recover this limit, uniformly expand a transparent pupil and divide its angular power by its area $A$. Its autocorrelation divided by $A$ tends to one at every fixed separation; equivalently, $H_{\mathscr{P}}/A$ tends to $\delta(\alpha)\delta(\beta)$ in the distributional sense. Equation~\eqref{supp:eq:angularconv} then gives the angular power per unit area and the outgoing radiance:
\begin{equation}\label{eq:transparent}
j_{\mathrm{out}}(\alpha,\beta)\equiv\lim_{A\to\infty}\frac1A\frac{\mathrm dP_A}{\mathrm d\Omega}=cL_{\mathrm{in}}(-\alpha,-\beta),\qquad
L_{\mathrm{out}}=\frac{j_{\mathrm{out}}}{c}=L_{\mathrm{in}}(-\alpha,-\beta).
\end{equation}
The radiance is preserved along each ray. The minus signs convert the source-facing coordinates into propagation coordinates. For the plane wave, the outgoing direction $\Omega_p$ has $(\alpha,\beta)=(-\alpha_0,-\beta_0)$ and $c=c_0$. Hence $j_{\mathrm{out}}=E_0\delta_\Omega(\Omega-\Omega_p)$ and $L_{\mathrm{out}}=(E_0/c_0)\delta_\Omega(\Omega-\Omega_p)$. The corresponding field is
\begin{equation*}
U(\xi,\eta,z)=\sqrt{E_0}\exp[-ik(\alpha_0\xi+\beta_0\eta)]\exp(ikc_0z).
\end{equation*}
Propagation changes only its phase; every plane parallel to the pupil retains irradiance $E_0$. These limits confirm the common normalization of incident irradiance, coherent response, and outgoing radiance.

\begin{figure}[!htbp]
\centering
\includegraphics[width=\linewidth]{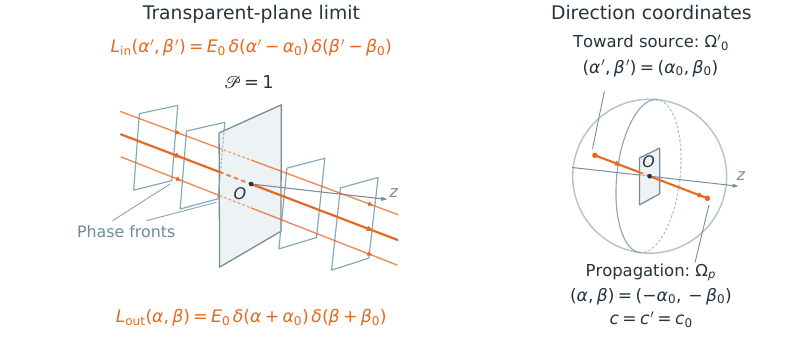}
\caption{Transparent-plane limit for a single plane wave. Source-facing coordinates $(\alpha',\beta')=(\alpha_0,\beta_0)$ and outgoing coordinates $(\alpha,\beta)=(-\alpha_0,-\beta_0)$ describe the same ray.}
\label{fig:supp_transparent_plane}
\end{figure}

\subsection{Relation to the Carter--Wolf Lambertian source}
\label{supp:carterwolf}
This comparison explains why the Lambertian source in main-text Section~\ref{sec:lambert} differs from the Carter--Wolf prescription: the two use different covariance and power conventions \cite{CarterWolf1975,CarterWolf1977}. Consider a statistically homogeneous planar source in the large-area limit, so its covariance depends only on separation and boundary effects vanish. For prescription $X$, write the covariance as $W_X(\Delta\mathbf r)$ and the radiant intensity per unit area as $j_X=\lim_{A\to\infty}A^{-1}\mathrm dP_X/\mathrm d\Omega$. In the present Fourier convention, the two prescriptions are separately
\begin{equation}\label{eq:CWpower}
\begin{aligned}
j_{\mathrm H}(\alpha,\beta)&=\frac{c}{\lambda^2}\widehat W_{\mathrm H}(\alpha/\lambda,\beta/\lambda),\\
j_{\mathrm{CW}}(\alpha,\beta)&=\frac{c^2}{\lambda^2}\widehat W_{\mathrm{CW}}(\alpha/\lambda,\beta/\lambda).
\end{aligned}
\end{equation}
Here $\mathrm H$ denotes the adopted Harvey plane-flux prescription and $\mathrm{CW}$ the Carter--Wolf ray-normal prescription. Equal emitted angular power, $j_{\mathrm H}=j_{\mathrm{CW}}$, requires
\begin{equation*}
\widehat W_{\mathrm{CW}}(\alpha/\lambda,\beta/\lambda)
=\frac{\widehat W_{\mathrm H}(\alpha/\lambda,\beta/\lambda)}{c},
\qquad (\alpha,\beta)\in\mathcal D.
\end{equation*}
For constant emitted radiance $L_0$, this condition is met by
\begin{equation}\label{eq:CWcovariance}
W_{\mathrm H}(\Delta\mathbf r)=\lambda^2L_0\delta^{(2)}(\Delta\mathbf r),\qquad
\widehat W_{\mathrm{CW}}(\alpha/\lambda,\beta/\lambda)=\frac{\lambda^2L_0}{c},\quad (\alpha,\beta)\in\mathcal D.
\end{equation}
The first spectrum is constant. Inverting the propagating part of the second gives, for $\rho=|\Delta\mathbf r|$,
\begin{align}\label{eq:CWsinc}
W_{\mathrm{CW,prop}}(\rho)&=L_0\iint_{\mathcal D}\frac{e^{ik(\alpha\Delta\xi+\beta\Delta\eta)}}{\sqrt{1-\alpha^2-\beta^2}}\,\mathrm d\alpha\,\mathrm d\beta\nonumber\\
&=2\pi L_0\int_0^1\frac{u\mathrm J_0(ku\rho)}{\sqrt{1-u^2}}\,\mathrm du=2\pi L_0\frac{\sin(k\rho)}{k\rho}.
\end{align}
Here $\mathrm{J}_0$ is the Bessel function. The delta covariance
with the single cosine and the sinc covariance with the squared
cosine both give $j=L_0\cos\theta$ and hemispherical exitance
$\pi L_0$ in the homogeneous-source limit. Their different
covariances can produce different finite-pupil predictions.
We use the single-cosine formulation as a scalar approximation
for unpolarized illumination, connecting delta-correlated source
statistics, radiance, and a reusable coherent response.
This approximation is motivated by the vector-aperture benchmark for unpolarized illumination in Section \ref{supp:angular_power} and by the experimental scattering comparisons discussed in Section \ref{sec:angular_power} of the main article.

\subsection{Matching a reusable response to the central emitter}
\label{supp:central_response}
The correction in main-text Section~\ref{sec:curvature} tests how much of the oblique-incidence error can be removed while retaining one reusable response. We match that response to the central emitter, then use the same angular convolution. To identify the correction, expand the source-to-pupil distance about the pupil center, with $\mathbf r=(\xi,\eta)$:
\begin{equation}\label{eq:pathexpand}
d=R'-\alpha'\xi-\beta'\eta+\frac{\xi^2+\eta^2-(\alpha'\xi+\beta'\eta)^2}{2R'}+O(|\mathbf r|^3/R'^2).
\end{equation}
For the central direction, let $r_\parallel$ and $r_\perp$ denote pupil coordinates parallel and perpendicular to its transverse projection. The quadratic phase is $\phi_2=\pi(c_0^3r_\parallel^2+c_0r_\perp^2)/(\lambda z_1)$, where $c_0=\cos\theta_0$. The common on-axis phase replaces both cosine factors by one, giving the excess
\begin{equation}\label{eq:diagnostic}
\Delta\phi_2=\frac\pi{\lambda z_1}\left[(1-c_0^3)r_\parallel^2+(1-c_0)r_\perp^2\right],\qquad
\chi_0=\pi F_1[1-\cos^3\theta_0],\qquad F_1=\frac{a^2}{\lambda z_1}.
\end{equation}
Here $a$ is the radius of an enclosing disk, and $\chi_0$ is the largest quadratic-phase mismatch over that disk. The intensity error also depends on pupil weighting, source width, amplitude variation, and higher-order phase terms. At fixed $F_1$ and $\theta_0$, amplitude variation also depends on $a/\lambda$.

Choose the central emitter at $(x'_0,y'_0,-z_1)$, and define $\mathbf r'_0=(x'_0,y'_0)$, $R'_0=\sqrt{z_1^2+|\mathbf r'_0|^2}$, $\mathbf q_0=(\alpha_0,\beta_0)=\mathbf r'_0/R'_0$, and $d_0(\mathbf r)=\sqrt{z_1^2+|\mathbf r-\mathbf r'_0|^2}$. Matching its quadratic phase gives
\begin{equation}\label{eq:Qquad}
Q_{\mathrm{quad}}(\xi,\eta)=\mathscr{P}(\xi,\eta)\exp\!\left\{\frac{i\pi}{\lambda R'_0}\left[(1-\alpha_0^2)\xi^2-2\alpha_0\beta_0\xi\eta+(1-\beta_0^2)\eta^2\right]\right\}.
\end{equation}
Retaining its full incident phase and amplitude instead gives
\begin{equation}\label{eq:Qcentral}
Q_{\mathrm{central}}(\mathbf r)=\mathscr{P}(\mathbf r)\left(\frac{R'_0}{d_0}\right)^2\left(1+\frac{i}{kd_0}\right)e^{ik[d_0-R'_0+\mathbf q_0\cdot\mathbf r]}.
\end{equation}
The positive linear phase removes the tilt already supplied by the angular source map. Either expression replaces the common-curvature pupil in $H_Q$, leaving $L_{\mathrm{in}}$ and Eq.~\eqref{supp:eq:angularconv} unchanged. The full correction reproduces the selected emitter within the incident scalar kernel; using it for neighboring emitters still approximates their different wavefronts.

\begin{figure}[!htbp]
\centering
\includegraphics[width=\linewidth]{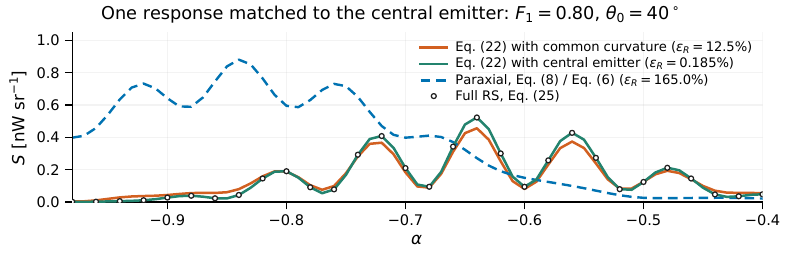}
\caption{Central-emitter correction for the double slit at $F_1=0.80$ and $\theta_0=40^\circ$. Matching the central wavefront improves the $\beta=0$ profile while retaining one reusable response. Legend errors are two-dimensional $\epsilon_R$; the source, pupil, and absolute scale are unchanged.}
\label{fig:curvature_correction}
\end{figure}

For the double slit at $F_1=0.80$, $\theta_0=40^\circ$, and $\sigma_t=0.025$, the angular error $\epsilon_R$ falls from 12.5\% with common on-axis curvature to 3.86\% with Eq.~\eqref{eq:Qquad} and 0.185\% with Eq.~\eqref{eq:Qcentral} (Fig.~\ref{fig:curvature_correction}). The source, pupil, and radiance are unchanged. Further source and pupil quadrature refinement has negligible effect; halving the angular spacing changes these errors by less than 0.071\% relatively. With the full correction, the remaining error reflects wavefront variation across the source. The main validity maps and timings use the original common-curvature response.

\subsection{Finite-monitor remapping and its approximation}
\label{supp:monitor_remap}
Main-text Eq.~\eqref{eq:monitorconv} remaps one coherent monitor response in direction cosines. This introduces an outgoing-propagation approximation, separate from the incident-wavefront approximation above. To identify it, factor the full source-to-pupil-to-monitor integral, main-text Eq.~\eqref{eq:monitordirect}, as
\begin{equation}
\begin{aligned}
 A_{\alpha',\beta'}(\xi,\eta)&=\frac{R'^2}{R_1^2}
 \left(1+\frac{i}{kR_1}\right)e^{ik(R_1-R'+\alpha'\xi+\beta'\eta)},\\
 B_{x,y}(\xi,\eta)&=\left(\frac{\rho}{R}\right)^2
 \left(1+\frac{i}{kR}\right)e^{ik(R-\rho+\alpha\xi+\beta\eta)},\\
 K_{x,y;\alpha',\beta'}(u,v)&=\frac1{\lambda^2}
 \left|\mathcal F_{\xi,\eta}\{\mathscr P A_{\alpha',\beta'}B_{x,y}\}\!\left(\frac{u}{\lambda},\frac{v}{\lambda}\right)\right|^2,\\
 E(x,y)&=\frac{c^4}{z_2^2}\iint_{\mathcal D}L_{\mathrm{in}}(\alpha',\beta')
 K_{x,y;\alpha',\beta'}(\alpha+\alpha',\beta+\beta')\,\mathrm d\alpha'\,\mathrm d\beta'.
\end{aligned}
\label{eq:monitor_kernel}
\end{equation}
Here $R_1$, $R$, and $R'$ are the distances defined in main-text Eq.~\eqref{eq:monitordirect}, $\rho=\sqrt{z_2^2+x^2+y^2}$, $(\alpha,\beta)=(x,y)/\rho$, and $c=z_2/\rho$. The factors $A_{\alpha',\beta'}$ and $B_{x,y}$ retain the incident and outgoing amplitude and phase after removing their central distances and linear tilts. This factorization makes no propagation approximation.

Now replace the incident pupil $\mathscr P A_{\alpha',\beta'}$ by the common reference $Q_{z_1}$, and write $\mathbf x=(x,y)$, $\mathbf q=(\alpha,\beta)$, $\mathbf p=(\alpha',\beta')$, and $\mathbf s=\mathbf q+\mathbf p$. Define $K^Q_{\mathbf x}(\mathbf s)=\lambda^{-2}|\mathcal F\{Q_{z_1}B_{\mathbf x}\}(\mathbf s/\lambda)|^2$. For $|\mathbf s|<1$, let $c_s=\sqrt{1-|\mathbf s|^2}$ and $\mathbf x_s=z_2\mathbf s/c_s$. The reference monitor response satisfies
\begin{equation*}
\frac{z_2^2}{c_s^4}H^{\mathrm{mon}}_{z_1,z_2}(\mathbf x_s)=K^Q_{\mathbf x_s}(\mathbf s).
\end{equation*}
The required response is $K^Q_{\mathbf x}(\mathbf s)$, whereas Eq.~\eqref{eq:monitorconv} uses $K^Q_{\mathbf x_s}(\mathbf s)$: remapping evaluates the outgoing factor at $\mathbf x_s$ instead of $\mathbf x$. For axial incidence, $\mathbf p=0$ and these points coincide, making this outgoing step exact for the prescribed $Q_{z_1}$. In the outgoing far field, both $B$ factors tend to one and angular convolution is recovered. The Fresnel/paraxial limit recovers GVS, with a residual pupil phase independent of monitor position.

The leading outgoing phase is quadratic in $\mathbf r$, with curvature matrix $c(I-\mathbf q\mathbf q^{\mathrm T})/z_2$. Comparing this curvature at $\mathbf x$ and $\mathbf x_s$ gives the largest quadratic-phase mismatch over a disk of radius $a$:
\begin{equation*}
\chi_2^{(2)}=\pi F_2\left\|
c_s(I-\mathbf s\mathbf s^{\mathrm T})
-c(I-\mathbf q\mathbf q^{\mathrm T})\right\|_2,
\qquad F_2=\frac{a^2}{\lambda z_2}.
\end{equation*}
The spectral norm includes both transverse curvatures and their orientation. At the undeviated ray's monitor intercept, $\mathbf q=-\mathbf p$, this reduces to $\chi_2^{(2)}=\pi F_2(1-c'^3)$. The remapping error can therefore remain even for a single oblique emitter. Like $\chi_0$, this quantity measures phase mismatch rather than bounding the intensity error.

As $|\mathbf s|\to1^-$, $\mathbf x_s$ recedes to infinity and $B_{\mathbf x_s}\to1$, so the remapped response approaches $H_{z_1}(\mathbf s)$. Equation~\eqref{eq:monitorconv} uses this response beyond the shifted unit disk; the quadratic diagnostic applies inside it.
\endgroup

\section{Reference calculations and error measures}
\label{supp:checks}
The comparisons in main-text Section~\ref{sec:validation} separate three questions: whether the formulas are algebraically equivalent, whether their power scales agree, and how much error their propagation approximations introduce.

\subsection{Angular and finite-monitor references}
To test the common incident curvature, the angular reference retains each emitter's full source-to-pupil Rayleigh--Sommerfeld (RS) kernel:
\begin{equation}\label{eq:fullkernel}
h_{z_1}(\mathbf r,\mathbf r')=\frac{z_1}{i\lambda d^2}\left(1+\frac{i}{kd}\right)e^{ikd},\qquad
d=\sqrt{z_1^2+|\mathbf r-\mathbf r'|^2}.
\end{equation}
We calculate the coherent pupil response for each emitter and then add intensities:
\begin{equation}\label{eq:fullreference}
S_{\mathrm{ref}}(\alpha,\beta)=\frac1{\lambda^2}\iint\lambda^2L_0(\mathbf r')
\left|\iint\mathscr P(\mathbf r)h_{z_1}(\mathbf r,\mathbf r')
e^{-ik(\alpha\xi+\beta\eta)}\,\mathrm d^2\mathbf r\right|^2\,\mathrm d^2\mathbf r'.
\end{equation}
This retains the source-dependent phase, amplitude, obliquity, and $1/(kd)$ correction. Both predictions use $R=cS$, so their difference tests the incident approximation.

For a finite monitor, the reference instead retains both propagation legs:
\begin{equation}\label{eq:monitorreference}
E_{\mathrm{ref}}(\mathbf x)=\iint\lambda^2L_0(\mathbf r')
\left|\iint\mathscr P(\mathbf r)h_{z_1}(\mathbf r,\mathbf r')
h_{z_2}(\mathbf x,\mathbf r)\,\mathrm d^2\mathbf r\right|^2\,\mathrm d^2\mathbf r'.
\end{equation}
Here $h_{z_2}$ has the form of Eq.~\eqref{eq:fullkernel} with the pupil and monitor coordinates. This is main-text Eq.~\eqref{eq:monitordirect}; it tests the incident approximation and outgoing remapping together. The VCSEL example prescribes a collimated angular mixture, so only the outgoing RS propagation is required there.

\subsection{Errors on the absolute irradiance scale}
We compare full distributions without fitting an amplitude or shifting either prediction. The angular and monitor errors are
\begin{align}
\epsilon_R&=\left[\frac{\iint_{\mathcal R}c|S-S_{\mathrm{ref}}|^2\,\mathrm d\alpha\,\mathrm d\beta}
{\iint_{\mathcal R}c|S_{\mathrm{ref}}|^2\,\mathrm d\alpha\,\mathrm d\beta}\right]^{1/2},\label{eq:angular_error}\\
\epsilon_E&=\left[\frac{\int_{\mathcal A}|E-E_{\mathrm{ref}}|^2\,\mathrm dA}
{\int_{\mathcal A}|E_{\mathrm{ref}}|^2\,\mathrm dA}\right]^{1/2}.\label{eq:monitor_error}
\end{align}
The angular weight follows from $R=cS$ and $\mathrm d\Omega=\mathrm d\alpha\,\mathrm d\beta/c$, giving the radiant-intensity error in main-text Eq.~\eqref{eq:error}. Pointwise checks instead divide $|E-E_{\mathrm{ref}}|$ by the reference image maximum; these are reported separately from the $L^2$ errors.

Both models use the same integration region. In the angular study, it includes the nonparaxial and paraxial image centers, $-\mathbf t_0/\sqrt{1+|\mathbf t_0|^2}$ and $-\mathbf t_0$, with margins 0.26 horizontally and 0.24 vertically. Bounds are clipped to $[-0.98,0.98]$, and points outside the propagating disk are excluded. Numerical refinement measures changes in a calculation; model discrepancy measures differences between propagation formulas.

\subsection{Algebraic and normalization checks}
Three independent implementations test the Fresnel equivalence in Section~\ref{supp:proof}: the nested transform with direct linear pupil autocorrelation, the convolution in mapped screen coordinates, and propagation of each source field through two Fresnel kernels followed by intensity summation. Unequal off-axis source spots, an asymmetric complex pupil, and an offset rectangular grid expose inversion and scaling errors. Six cases cover $M=0.5,1,2$ at wavelengths 532 and 940~nm.

With identical integration weights, all three implementations agree to floating-point precision. Independent pupil refinement changes irradiance by less than $0.05\%$; source refinement is negligible at that scale. Omitting the source inversion or its area factor breaks the agreement. These checks distinguish the algebraic identity from the accuracy of its numerical evaluation.

Independent normalization checks recover the forward Lambertian power $\pi\iint L_0\,\mathrm dA'$, match source-plane and angular irradiance integrals, and reproduce $E=(P_s/\pi)z_1^2/R_s'^4$ for a localized emitter. The formulations therefore share the same absolute power scale.

\section{Numerical convergence and scope of the validity study}
\label{supp:convergence}
\subsection{Separating numerical and model error}
The validity map in main-text Fig.~\ref{fig:maps} should measure the common-curvature approximation rather than insufficient numerical resolution. At infinite source distance, that approximation disappears: angular convolution and direct source summation agree to floating-point precision on identical pupil samples. An independently integrated continuous pupil leaves a discrepancy below $0.003\%$, identifying the rasterization error in this test.

The finite-distance map contains 396 cases. We refine pupil pitch, angular spacing, source and pupil quadrature, padding, and integration region separately \cite{ShenWang2006,VoelzRoggemann2009,MatsushimaShimobaba2009}. Three representative cases cover the double slit at $F_1=0.05$ and 0.80 and the circular pupil at $F_1=0.20$, all at $\theta_0=40^\circ$. In each case, the largest numerical change is below one tenth of the model discrepancy. The $12.5\%$ error at $F_1=0.80$ therefore reflects the propagation approximation.

\subsection{Source width and pupil shape}
The central phase mismatch $\chi_0$ alone cannot predict intensity error, because source averaging and pupil shape also change the diffraction pattern. At fixed $F_1=0.20$, $\theta_0=40^\circ$, and $\chi_0=0.3459$~rad, increasing $\sigma_t$ from 0.0125 to 0.05 reduces the circular-pupil error from $2.024\%$ to $1.660\%$; the double slit shows the same trend. Both pupils have enclosing radius $a=8.3596~\mu$m and $a/\lambda=8.36$.

A second test keeps the physical source size, radiance, and lateral offset fixed while reducing its distance. Error increases as the source approaches; here the Fresnel number, incidence angle, and angular width change together. A smooth, nonseparable complex pupil under two-axis incidence shows the same increase with $F_1$, reaching $10.9\%$ at $F_1=0.80$. The finite-source limitation therefore extends beyond binary apertures.

\subsection{Finite-monitor double-slit comparison}
\label{supp:finite_slit}
Main-text Fig.~\ref{fig:profiles}(k--m) tests the incident approximation and monitor remapping together. It retains the Gaussian source and double slit, with $F_1=F_2=0.2$ ($z_1=z_2=0.349$~mm), and compares Eq.~\eqref{eq:monitorconv} with the two-leg RS reference.

Both convolution models use the same RS-propagated coherent response. With $t_x=x^{\prime} / z_1$ and $t_y=y^{\prime} / z_1$, Eq.~\eqref{eq:psfconv} translates it across the physical screen:
\begin{equation}
E_8(x,y)=\iint_{\mathbb R^2}
L_0(z_1t_x,z_1t_y)\,
H^{\mathrm{mon}}_{z_1,z_2}(x+z_2t_x,y+z_2t_y)
\,\mathrm dt_x\,\mathrm dt_y.
\label{eq:monitor_paraxial_comparator}
\end{equation}
Equation~\eqref{eq:monitorconv} instead combines the responses
in direction-cosine coordinates. For this model and the RS reference,
we use $\alpha'=t_x/\sqrt{1+t_x^2+t_y^2}$ and
$\beta'=t_y/\sqrt{1+t_x^2+t_y^2}$, with
$\mathrm d\alpha'\,\mathrm d\beta'
=(1+t_x^2+t_y^2)^{-2}\,\mathrm dt_x\,\mathrm dt_y$.
All predictions retain their absolute irradiance scale.

We integrate over the slit openings to compute the coherent response
and reference profiles, and check selected monitor positions with
finer source, pupil, and convolution sampling. Over a fixed
two-dimensional monitor region, the relative $L^2$ irradiance error
of Eq.~\eqref{eq:monitorconv} against the RS reference remains
below $1.8\%$ at each tested incidence angle:
$0^\circ$, $20^\circ$, and $40^\circ$.

\section{Finite-monitor $3\times3$ phase grating}
\label{supp:grating}
\subsection{Fixed grating and propagation models}
\begin{figure}[!htbp]
\centering
\includegraphics[width=\linewidth]{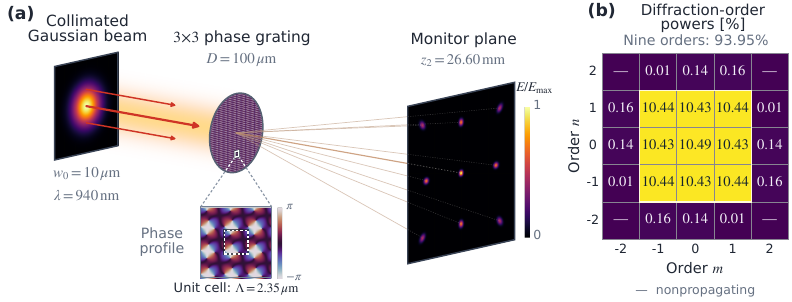}
\caption{Coherent Gaussian illumination of the fixed grating. (a) Source and angular-spectrum monitor irradiance, each normalized to its own peak. The inset shows three phase periods per axis and outlines one cell; geometry is schematic. (b) Incident-power percentages $100|b_{mn}|^2$; dashes mark nonpropagating orders.}
\label{fig:supp_grating_gaussian}
\end{figure}

We use the same grating in all three models to compare image distortion and irradiance variation. Figure~\ref{fig:supp_grating_gaussian} shows how it splits a collimated Gaussian beam into nine main orders. With extended incoherent illumination, each order forms a source image (main-text Section~\ref{sec:grating}). The pupil transmission is
\begin{equation}
\mathscr P(\xi,\eta)=\mathbf1_{\xi^2+\eta^2\le a^2}
\exp\!\left[i\phi\!\left(\frac{\xi}{\Lambda}\bmod1,\frac{\eta}{\Lambda}\bmod1\right)\right],
\qquad a=50~\mu\mathrm m,\quad\Lambda=2.35~\mu\mathrm m.\label{eq:grating_pupil}
\end{equation}
The optimized $512^2$ phase cell repeats across the pupil. Its ideal order amplitudes and power fractions are
\begin{equation}
b_{mn}=\frac1{\Lambda^2}\int_0^\Lambda\!\int_0^\Lambda
e^{i\phi(\xi/\Lambda,\eta/\Lambda)}
e^{-i2\pi(m\xi+n\eta)/\Lambda}\,\mathrm d\xi\,\mathrm d\eta,
\qquad I_{mn}=|b_{mn}|^2.\label{eq:grating_orders}
\end{equation}
The nine target orders carry $\eta_{\rm cell}=\sum_{m,n=-1}^{1}I_{mn}=93.95\%$ of the incident power, with peak-to-valley nonuniformity $U_{\rm cell}=(I_{\max}-I_{\min})/(I_{\max}+I_{\min})=0.32\%$. These metrics characterize the periodic cell. The image comparisons retain the full phase, all harmonics, and their interference on an absolute irradiance scale.

Both convolution models use the same axial coherent monitor response. Equation~\eqref{eq:psfconv} shifts it in monitor coordinates; Eq.~\eqref{eq:monitorconv} maps it into direction cosines before source averaging. The RS reference, Eq.~\eqref{eq:monitordirect}, retains each emitter's incident and outgoing kernels in Eq.~\eqref{eq:monitor_kernel}. At $F_1=0.00887$, the Eq.~\eqref{eq:monitorconv}--RS comparison therefore tests both incident curvature and monitor remapping.
\FloatBarrier

\subsection{Image distortion and irradiance variation}
\begin{figure}[!htbp]\centering
\includegraphics[width=\linewidth]{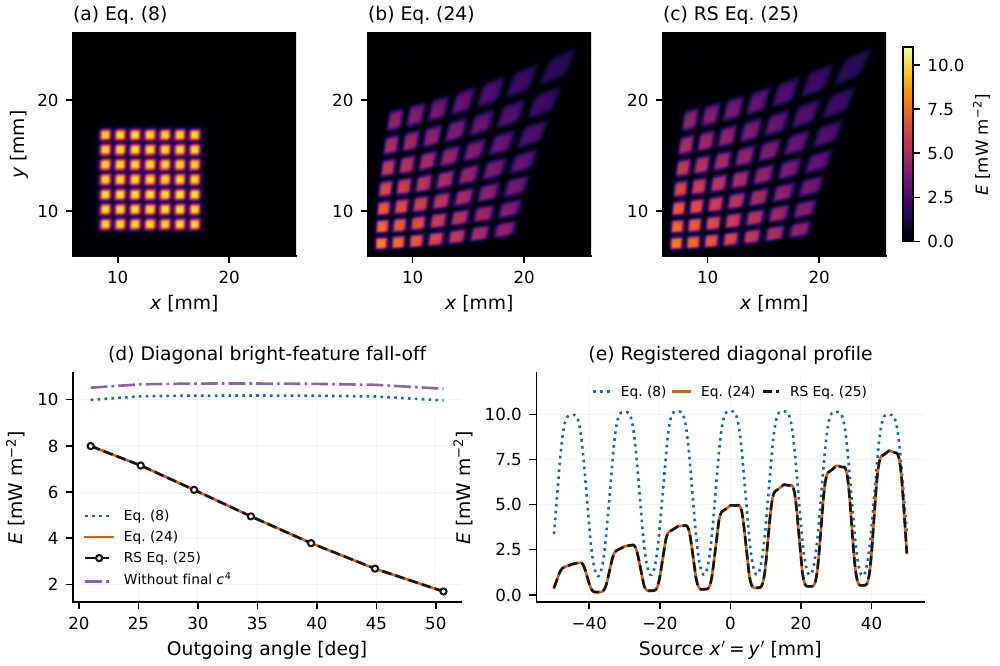}
\caption{Finite-monitor grating comparison. (a--c) Diagonal-order details on a common absolute irradiance scale. (d) Irradiance at the predicted locations of the same emitting squares; the additional curve omits only the final $c^4$ monitor factor. (e) Profiles through corresponding bright squares and dark gaps.}
\label{fig:grating_diagnostics}
\end{figure}

The square source array makes the distortion visible: nonzero orders stretch, while the zero-order chief-ray map remains $\mathbf x=-z_2\mathbf r'/z_1$. To compare corresponding bright features, Fig.~\ref{fig:grating_diagnostics} samples each model at its predicted locations of the same emitting squares. This separates irradiance variation from differences in image position.

The RS bright-feature irradiance decreases by $44\%$ along a side order and $79\%$ along a diagonal order. Comparing corresponding features makes clear that this decrease accompanies the geometric distortion.

Dividing the Eq.~\eqref{eq:monitorconv} image by its final $c^4$ monitor factor isolates the contribution of this projection weight. The angular illumination, coherent kernel, and mapped feature positions remain fixed. Figure~\ref{fig:grating_diagnostics}(d) shows how this change affects the predicted irradiance decrease.

\subsection{Numerical resolution and reference checks}
The checks behind main-text Fig.~\ref{fig:grating} establish the numerical scale of the image comparison. For the square source, the whole-image Eq.~\eqref{eq:monitorconv}--RS discrepancy is $0.046\%$ in relative $L^2$. Refining source integration, pupil sampling, the angular grid, and RS-kernel interpolation separately changes the image by less than $0.05\%$ in the same norm. The reported model discrepancy is therefore comparable to the largest numerical change.

Independent direct integration at selected bright features, gaps, order centers, and corners agrees with the image reference to within $0.08\%$ of its peak. This pointwise check also covers nearly dark pixels without dividing by their local irradiance.

For the grayscale USAF source, the Eq.~\eqref{eq:monitorconv}--RS discrepancy is $0.031\%$ in relative $L^2$; subdividing each source pixel changes the image by $0.011\%$. Both source patterns therefore show the same distortion and irradiance fall-off, with Eq.~\eqref{eq:monitorconv} and RS agreeing closely at the tested numerical resolution.

\FloatBarrier
\section{VCSEL-array diffuser: design and validation}
\label{supp:vcsel}

\subsection{Illumination and phase representation}
Main-text Fig.~\ref{fig:vcsel} tests whether a diffuser optimized for uniform wide-angle illumination achieves that uniformity under RS propagation. Four models optimize the same target for a collimated VCSEL array.

The illumination consists of 49 mutually incoherent emitters with $40~\mu$m pitch, $12~\mu$m near-field intensity FWHM, and $5$~mW per emitter at $\lambda=940$~nm. The source and diffuser lie in the front and back focal planes of an ideal $f=1.5$~mm sine-condition collimator. Each emitter maps to a Gaussian angular spot of FWHM $12~\mu\mathrm m/f$. All emitters share a Gaussian envelope on the diffuser with $1/e^2$ intensity radius $w=264.49~\mu$m. Angular components within each spot are also treated as mutually incoherent. This prescribed collimated mixture requires no finite-source curvature approximation.

We optimize a real, unwrapped phase variable in waves on a $2048^2$ grid covering a $128~\mu$m periodic cell (62.5~nm pitch). A periodic Gaussian filter with standard deviation $0.1~\mu$m smooths the variable before conversion to radians. All models use this parameterization. Device-specific emitter modes, polarization, material transmission, and collimator aberrations lie outside this scalar phase-screen model.

\subsection{Target, optimization, and RS evaluation}
The objective rewards uniform illumination within the desired field while limiting spill outside it. At $z_2=1$~m, the nominal field is $|x|\le z_2\tan50^\circ$, $|y|\le z_2\tan35^\circ$. The target has a nearly uniform interior and an edge roll-off set by the finite source; its residual interior nonuniformity is approximately $0.1\%$. The interior mask $I$ contains points whose source footprint remains inside the field. The remaining illuminated edge region is $G$, and the dark exterior is $O$. With normalized target $T$ and interior mean $s=\langle E\rangle_I$, we minimize
\begin{equation}
\mathcal L=
\langle(E/s-T)^2\rangle_I+
\langle(E/s-T)^2\rangle_G+
0.3\langle(E/s)^2\rangle_O+
0.02(1-\eta_{\rm FoI}).\label{eq:vcsel_objective}
\end{equation}
The first two terms match the target shape, the third suppresses exterior light, and the last rewards collected power. Here
$\eta_{\rm FoI}=P_0^{-1}\int_{\rm FoI}E\,\mathrm dA$
is the fraction of incident power collected within the field. We report interior uniformity as
$U_{\rm RMS}=\operatorname{std}_I(E)/\langle E\rangle_I$.

The coordinate-corrected Eq.~\eqref{eq:psfconv} comparator omits the $c^4$ factor from its irradiance-shape model but retains it when integrating power. It also continues to translate responses in monitor coordinates. Section~\ref{supp:grating} separately isolates the effect of the projection weight alone.

The four models are optimized separately from the same three initial phase arrays, formed by adding smooth random perturbations to a diverging lenslet. Each run uses 800 Adam iterations, JAX autodifferentiation, double-precision FFTs, and a cosine phase-step schedule from 0.08 to 0.008~rad. Across the three starts, both nonparaxial design families have standard deviations below 0.04 percentage points in uniformity and efficiency when evaluated with Eq.~\eqref{eq:monitorconv}. Table~\ref{tab:vcsel_rs} reports the independent RS evaluation of the first matched start.

\begin{table}[!htbp]\centering
\caption{Common RS evaluation of the first matched VCSEL start at $z_2=1$~m. Uniformity is measured over the interior; efficiency is relative to incident power.}
\label{tab:vcsel_rs}
\begin{tabular}{@{}lrr@{}}\toprule
Optimization model & RS $U_{\rm RMS}$ [\%] & RS $\eta_{\rm FoI}$ [\%]\\\midrule
Paraxial Eq.~\eqref{eq:psfconv} & 32.287 & 47.259\\
Coordinate-corrected Eq.~\eqref{eq:psfconv} & 33.904 & 92.150\\
Far-field Eq.~\eqref{eq:common} & 1.272 & 84.210\\
Finite-distance Eq.~\eqref{eq:monitorconv} & 1.272 & 84.210\\\bottomrule
\end{tabular}
\end{table}

\subsection{Finite-distance response and validation}
The finite-distance correction explains why Eqs.~\eqref{eq:common} and \eqref{eq:monitorconv} give nearly the same design in this geometry. Gaussian illumination gives each diffraction order a narrow Gaussian lobe. Mapping its monitor response into direction cosines adds radial and tangential variances $(c_m^3w/2z_2)^2$ and $(c_mw/2z_2)^2$, where $c_m=\sqrt{1-|\mathbf u_m|^2}$ for order direction $\mathbf u_m$. The optimizer includes this small broadening to first order in the Fourier domain. Direct Gaussian-lobe calculations and the mapped full RS response check the approximation.

The whole-image RS reference propagates the prescribed collimated mixture through the outgoing kernel of Eq.~\eqref{eq:monitordirect}, retaining interference between orders with a bounded truncation of the order-pair sum. For the Eq.~\eqref{eq:monitorconv} design, the relative $L^2$ discrepancies from RS are $0.0479\%$ for Eq.~\eqref{eq:common} and $0.0493\%$ for Eq.~\eqref{eq:monitorconv}. The shared Eq.~\eqref{eq:common}/Eq.~\eqref{eq:monitorconv} design label in main-text Fig.~\ref{fig:vcsel} reflects this agreement; the displayed pupil is evaluated with Eq.~\eqref{eq:monitordirect}.

Pupil, angular, and detector refinements keep the checked nonparaxial design below $1.3\%$ RMS nonuniformity; the largest efficiency change is approximately 0.3 percentage points. Detailed profile and focal-length checks accompany the numerical records. At $F_2=0.074$, both design models thus reproduce the RS image to about $0.05\%$ in relative $L^2$.

\FloatBarrier
\section{Finite-distance design of irradiance patterns}
\label{supp:grating_inverse}
\subsection{Flat-top array and design objective}
Main-text Fig.~\ref{fig:grating_inverse} tests whether finite-distance optimization improves the shape and uniformity of nine illuminated squares beyond paraxial refocusing of a far-field design. The five extended Lambertian emitters are those specified in main-text Section~\ref{sec:grating_inverse}. We use $\lambda=940$~nm, pupil radius $a=50~\mu$m, $z_1=300$~mm, and $z_2=a^2/\lambda=2.6596$~mm ($F_2=1$).

The target centers are $\mathbf x_{mn}=z_2\mathbf q_{mn}/\sqrt{1-|\mathbf q_{mn}|^2}$, with $\mathbf q_{mn}=0.4(m,n)$ and $m,n\in\{-1,0,1\}$. Each target square is $320~\mu$m wide, with edges smoothed by a Gaussian of standard deviation $20~\mu$m. The dimensionless target $T$ approaches one in the interiors. We evaluate it on nine $420\times420~\mu\mathrm m$ tiles, whose union is $\mathcal A$. Let $I=\{T>0.95\}$ denote their combined interiors, $C=\{T\geq0.5\}$ the collection region, and $\overline E_I=\langle E\rangle_I$. The objective is
\begin{equation}
\mathcal L=\left\langle\left(E/\overline E_I-T\right)^2\right\rangle_{\mathcal A}
+0.05(1-\eta_{\rm sq}),\qquad
\eta_{\rm sq}=\frac{1}{P_0}\int_C E\,\mathrm dA.\label{eq:flattop_objective}
\end{equation}
The first term matches the irradiance profile, including its edges and dark surroundings; the second rewards collection relative to incident pupil power $P_0$. We report
\begin{equation}
\epsilon_{\rm shape}=\frac{\|E/\overline E_I-T\|_{L^2(\mathcal A)}}{\|T\|_{L^2(\mathcal A)}},
\qquad U=\frac{\operatorname{std}_I(E)}{\overline E_I}.\label{eq:flattop_metrics}
\end{equation}
Shape error measures agreement with the full target, while $U$ measures spatial variation across all nine interiors. Both use one common interior mean for each design.

\subsection{Phase optimization with one reusable response}
We optimize an independent phase on each illuminated pixel of a $213\times213$ grid with $0.47~\mu$m pitch. This free-form pupil can change the spot profiles as well as their collected powers. Equations~\eqref{eq:common} and \eqref{eq:monitorconv} start from the same sampled grating with a diverging quadratic phase and use the same objective and 2000 Adam iterations. The refocused comparator adds $-\pi(\xi^2+\eta^2)/(\lambda z_2)$ to the optimized Eq.~\eqref{eq:common} phase, with its strength fixed by the monitor distance.

The source and monitor sampling remain fixed during optimization. Their direction-cosine mapping, radiometric weights, and interpolation are therefore prepared once. Each trial phase supplies one coherent pupil response, from which the fixed maps form the incoherent irradiance. Source quadrature uses 120 samples across the five disks, weighted by $c'^4\,\mathrm dA'/z_1^2$ and normalized to unit pupil-center irradiance. The finite-distance response uses angular-spectrum integration over spectral patches covering the nine tiles; the notebook provides the sampling and implementation details.

\subsection{Common RS evaluation}
To check the patterns actually produced, we evaluate all three pupils by propagating each source sample separately and adding intensities. This reference retains each emitter's spherical incident phase and amplitude and uses the outgoing RS propagator. Table~\ref{tab:grating_inverse} reports the resulting metrics on the same monitor samples. At six positions covering axial, side, and diagonal squares, direct physical-pixel RS integration agrees with this spectral reference to within $0.005\%$ of each design's peak for all three pupils.

\begin{table}[!htbp]\centering
\caption{Source-resolved RS evaluation of the flat-top designs in main-text Fig.~\ref{fig:grating_inverse}. Shape error uses all nine tiles; spatial nonuniformity uses their combined interiors. Collection is relative to incident pupil power.}
\label{tab:grating_inverse}
\begin{tabular}{@{}lrrr@{}}\toprule
Design & Shape error [\%] & $U$ [\%] & Collection [\%]\\\midrule
Eq.~\eqref{eq:common} & 84.53 & 42.73 & 48.82\\
Eq.~\eqref{eq:common} + paraxial focus & 22.16 & 20.24 & 65.52\\
Eq.~\eqref{eq:monitorconv} & 4.90 & 4.50 & 63.74\\
\bottomrule
\end{tabular}
\end{table}

For the Eq.~\eqref{eq:monitorconv} design, its predicted irradiance differs from the source-resolved RS result by $0.27\%$ in relative $L^2$ norm over the nine tiles, without rescaling either prediction. This model comparison is distinct from the target shape error, which uses the interior-mean normalization above. Relative to nominal paraxial refocusing, finite-distance optimization reduces both shape error and spatial nonuniformity by about $4.5\times$, while the collected fraction falls from $65.52\%$ to $63.74\%$.

\clearpage
\subsection{Shaping a complex irradiance pattern}
\label{supp:complex_irradiance_pattern}
\begin{figure}[!htbp]
\centering
\includegraphics[width=\linewidth]{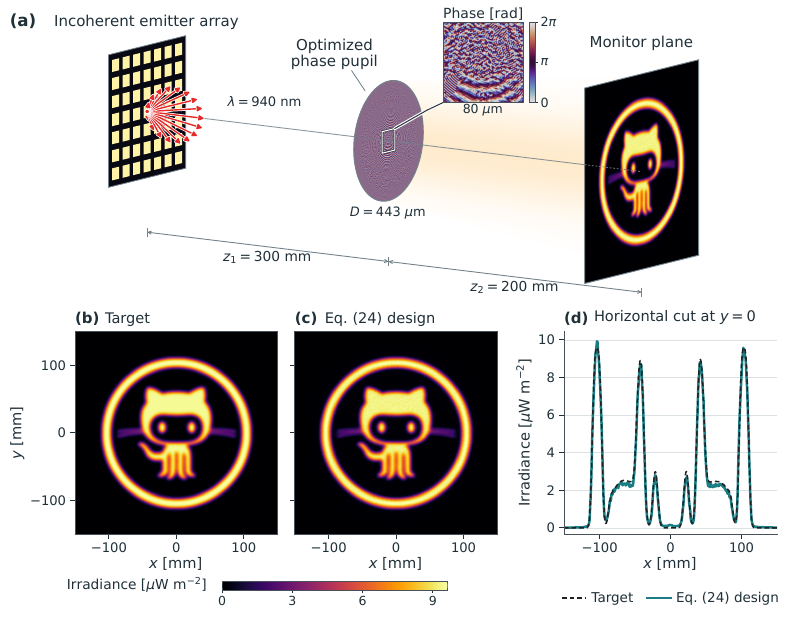}
\caption{Phase-only shaping of a grayscale image. (a) Emitter array, phase pupil, and monitor. The $7\times7$ source array of equally bright square emitters covers $8$~mm. (b) Target irradiance from a $220$~mm-wide grayscale logo. (c) RS irradiance [Eq.~\eqref{eq:monitordirect}] for the pupil optimized with Eq.~\eqref{eq:monitorconv}. (d) Target and RS irradiance profiles along $y=0$.}
\label{fig:supp_logo}
\end{figure}

The flat-top designs above are designed to give uniform irradiance across each square. To see if the same finite-distance convolution can also be used to reproduce variations in brightness within a single image, we take the grayscale logo in Fig.~\ref{fig:supp_logo} as our target. In main-text Fig.~\ref{fig:grating}, the grating forms copies of a USAF pattern already present in the source. Here the light comes from a $7\times7$ array of mutually incoherent square emitters ($0.80$~mm on a side, $1.20$~mm pitch). We scale the target irradiance $T$ so that its integral over the monitor equals the power transmitted by the pupil. We then optimize only the pupil phase over 4000 Adam steps, using the angular Fresnel response in Eq.~\eqref{eq:monitorconv}.

We independently evaluate the optimized pupil with the two-leg RS integral, Eq.~\eqref{eq:monitordirect}. Figure~\ref{fig:supp_logo}(c) shows the resulting irradiance. Over the full monitor, the relative target error $\|E-T\|_2/\|T\|_2$ is $3.36\%$ for this RS evaluation and $3.26\%$ for the prediction from Eq.~\eqref{eq:monitorconv}.

\clearpage

\FloatBarrier
\section{Timing protocol}
\label{supp:timing}
Main-text Fig.~\ref{fig:scaling} measures the speed of complete forward calculations at the same required accuracy, including the response construction needed when the pupil changes. The USAF benchmark uses $\lambda=940$~nm, a clear circular pupil of radius $a=16\lambda$, $F_1=0.02$, and $F_2=0.10$. A square monitor of width $512\lambda$ has $128^2$, $256^2$, or $512^2$ samples. The source spans $|x'|,|y'|\leq0.07z_1$; four sampling rules contain 128, 422, 1,207, and 3,553 nonzero weighted centroids. Both methods receive the same source samples and weights in each comparison.

The angular-spectrum (AS) method propagates each emitter's exact incident RS field with a band-limited angular spectrum and sums intensities before detector interpolation. Equation~\eqref{eq:monitorconv} uses one common-curvature response and an angular convolution. For each method, we select the fastest tested configuration that meets a $1\%$ relative $L^2$ irradiance-error requirement on the absolute scale.

A refined two-leg RS calculation supplies the reference for each source rule. Joint refinement and direct RS spot checks give an empirical reference allowance of approximately $0.004\%$, which is added to each candidate's image error before applying the $1\%$ threshold. The selected Eq.~\eqref{eq:monitorconv} and AS errors are approximately $0.17\%$ and $0.21\%$, respectively. Tested pitches, padding, and angular grids are recorded with the benchmark.

Timings use double precision, two FFT workers, and one BLAS thread. After one initial evaluation, we report the median of three complete calls. Each call constructs the source and propagation setup, evaluates the image, and maps it to the detector; Eq.~\eqref{eq:monitorconv} rebuilds its coherent response. Reference construction, error evaluation, gradients, and optimization are excluded. With 3,553 source samples, Eq.~\eqref{eq:monitorconv} is $164$--$174\times$ faster across the three monitor grids under this protocol.

\FloatBarrier
\bibliography{references}

@book{Goodman1985,
  author = {Goodman, Joseph W.},
  title = {{Statistical Optics}},
  publisher = {Wiley},
  year = {1985},
  url = {https://books.google.com/books/about/Statistical_Optics.html?id=2VTwAAAAMAAJ}
}

@article{VanCittert1934,
  author = {van Cittert, P. H.},
  title = {{Die wahrscheinliche Schwingungsverteilung in einer von einer Lichtquelle direkt oder mittels einer Linse beleuchteten Ebene}},
  journal = {Physica},
  year = {1934},
  volume = {1},
  pages = {201--210},
  doi = {10.1016/S0031-8914(34)90026-4},
  url = {https://doi.org/10.1016/S0031-8914(34)90026-4}
}

@article{Zernike1938,
  author = {Zernike, F.},
  title = {{The concept of degree of coherence and its application to optical problems}},
  journal = {Physica},
  year = {1938},
  volume = {5},
  number = {8},
  pages = {785--795},
  doi = {10.1016/S0031-8914(38)80203-2},
  url = {https://doi.org/10.1016/S0031-8914(38)80203-2}
}

@article{Hopkins1951,
  author = {Hopkins, H. H.},
  title = {{The concept of partial coherence in optics}},
  journal = {Proceedings of the Royal Society of London. Series A},
  year = {1951},
  volume = {208},
  number = {1093},
  pages = {263--277},
  doi = {10.1098/rspa.1951.0158},
  url = {https://doi.org/10.1098/rspa.1951.0158}
}

@article{MandelWolf1965,
  author = {Mandel, L. and Wolf, E.},
  title = {{Coherence Properties of Optical Fields}},
  journal = {Reviews of Modern Physics},
  year = {1965},
  volume = {37},
  pages = {231--287},
  doi = {10.1103/RevModPhys.37.231},
  url = {https://doi.org/10.1103/RevModPhys.37.231}
}

@article{Schell1967,
  author = {Schell, A. C.},
  title = {{A technique for the determination of the radiation pattern of a partially coherent aperture}},
  journal = {IEEE Transactions on Antennas and Propagation},
  year = {1967},
  volume = {15},
  number = {1},
  pages = {187--188},
  doi = {10.1109/TAP.1967.1138864},
  url = {https://doi.org/10.1109/TAP.1967.1138864}
}

@article{CarterWolf1977,
  author = {Carter, W. H. and Wolf, E.},
  title = {{Coherence and radiometry with quasihomogeneous planar sources}},
  journal = {Journal of the Optical Society of America},
  year = {1977},
  volume = {67},
  number = {6},
  pages = {785--796},
  doi = {10.1364/JOSA.67.000785},
  url = {https://doi.org/10.1364/JOSA.67.000785}
}

@article{Wolf1978,
  author = {Wolf, E.},
  title = {{Coherence and radiometry}},
  journal = {Journal of the Optical Society of America},
  year = {1978},
  volume = {68},
  number = {1},
  pages = {6--17},
  doi = {10.1364/JOSA.68.000006},
  url = {https://doi.org/10.1364/JOSA.68.000006}
}

@article{Nicodemus1963,
  author = {Nicodemus, F. E.},
  title = {{Radiance}},
  journal = {American Journal of Physics},
  year = {1963},
  volume = {31},
  number = {5},
  pages = {368--377},
  doi = {10.1119/1.1969512},
  url = {https://doi.org/10.1119/1.1969512}
}

@article{Smythe1947,
  author = {Smythe, W. R.},
  title = {{The Double Current Sheet in Diffraction}},
  journal = {Physical Review},
  year = {1947},
  volume = {72},
  number = {11},
  pages = {1066--1070},
  doi = {10.1103/PhysRev.72.1066},
  url = {https://doi.org/10.1103/PhysRev.72.1066}
}

@article{Walther1968,
  author = {Walther, A.},
  title = {{Radiometry and Coherence}},
  journal = {Journal of the Optical Society of America},
  year = {1968},
  volume = {58},
  number = {9},
  pages = {1256--1259},
  doi = {10.1364/JOSA.58.001256},
  url = {https://doi.org/10.1364/JOSA.58.001256}
}

@article{Nugent1990,
  author = {Nugent, K. A.},
  title = {{A generalization of Schell's theorem}},
  journal = {Optics Communications},
  year = {1990},
  volume = {79},
  number = {5},
  pages = {267--269},
  doi = {10.1016/0030-4018(90)90066-3},
  url = {https://doi.org/10.1016/0030-4018(90)90066-3}
}

@article{HarveyVernold1998,
  author = {Harvey, J. E. and Vernold, C. L.},
  title = {{Description of Diffraction Grating Behavior in Direction Cosine Space}},
  journal = {Applied Optics},
  year = {1998},
  volume = {37},
  number = {34},
  pages = {8158--8159},
  doi = {10.1364/AO.37.008158},
  url = {https://doi.org/10.1364/AO.37.008158}
}

@article{Harvey1999,
  author = {Harvey, J. E. and Vernold, C. L. and Krywonos, A. and Thompson, P. L.},
  title = {{Diffracted radiance: a fundamental quantity in nonparaxial scalar diffraction theory}},
  journal = {Applied Optics},
  year = {1999},
  volume = {38},
  number = {31},
  pages = {6469--6481},
  doi = {10.1364/AO.38.006469},
  url = {https://doi.org/10.1364/AO.38.006469}
}

@article{Montoya2025,
  author = {Montoya, Manuel and Lopera, Maria J. and Nie, Yunfeng and Blinder, David},
  title = {{Generalized Van Cittert-Zernike Schell propagator: an efficient algorithm for simulating partially coherent light}},
  journal = {Optics Express},
  year = {2025},
  volume = {33},
  number = {22},
  pages = {46313--46325},
  doi = {10.1364/OE.571673},
  url = {https://doi.org/10.1364/OE.571673}
}

@article{Muellers2022,
  author = {M{\"u}llers, David and Kuhl, Jonas and Kontermann, Stefan},
  title = {{Nonparaxial optical transfer function for arbitrary illumination in partially coherent imaging systems and the oblique source application}},
  journal = {Journal of the Optical Society of America A},
  year = {2022},
  volume = {39},
  number = {4},
  pages = {744--758},
  doi = {10.1364/JOSAA.452462},
  url = {https://doi.org/10.1364/JOSAA.452462}
}

@article{Bluestein1970,
  author = {Bluestein, L.},
  title = {{A linear filtering approach to the computation of discrete Fourier transform}},
  journal = {IEEE Transactions on Audio and Electroacoustics},
  year = {1970},
  volume = {18},
  number = {4},
  pages = {451--455},
  doi = {10.1109/TAU.1970.1162132},
  url = {https://doi.org/10.1109/TAU.1970.1162132}
}

@article{Montoya2026,
  author = {Montoya, Manuel and Nie, Yunfeng and Blinder, David},
  title = {{Computer-generated holography using the generalized Van Cittert-Zernike Schell propagator}},
  journal = {Optics Letters},
  year = {2026},
  volume = {51},
  number = {8},
  pages = {2084--2087},
  doi = {10.1364/OL.595741},
  url = {https://doi.org/10.1364/OL.595741}
}

@article{ShenWang2006,
  author = {Shen, Fabin and Wang, Anbo},
  title = {{Fast-Fourier-transform based numerical integration method for the Rayleigh-Sommerfeld diffraction formula}},
  journal = {Applied Optics},
  year = {2006},
  volume = {45},
  number = {6},
  pages = {1102--1110},
  doi = {10.1364/AO.45.001102},
  url = {https://doi.org/10.1364/AO.45.001102}
}

@article{VoelzRoggemann2009,
  author = {Voelz, David G. and Roggemann, Michael C.},
  title = {{Digital simulation of scalar optical diffraction: revisiting chirp function sampling criteria and consequences}},
  journal = {Applied Optics},
  year = {2009},
  volume = {48},
  number = {32},
  pages = {6132--6142},
  doi = {10.1364/AO.48.006132},
  url = {https://doi.org/10.1364/AO.48.006132}
}

@article{MatsushimaShimobaba2009,
  author = {Matsushima, Kyoji and Shimobaba, Tomoyoshi},
  title = {{Band-limited angular spectrum method for numerical simulation of free-space propagation in far and near fields}},
  journal = {Optics Express},
  year = {2009},
  volume = {17},
  number = {22},
  pages = {19662--19673},
  doi = {10.1364/OE.17.019662},
  url = {https://doi.org/10.1364/OE.17.019662}
}

@article{Harvey2000Errata,
  author = {Harvey, J. E. and Vernold, C. L. and Krywonos, A. and Thompson, P. L.},
  title = {{Diffracted radiance: a fundamental quantity in nonparaxial scalar diffraction theory: errata}},
  journal = {Applied Optics},
  year = {2000},
  volume = {39},
  number = {34},
  pages = {6374--6375},
  doi = {10.1364/AO.39.006374},
  url = {https://doi.org/10.1364/AO.39.006374}
}

@article{Zissa2018,
  author = {Zissa, David E.},
  title = {{Diffracted radiance not proven to be fundamental quantity in nonparaxial scalar diffraction theory: comment}},
  journal = {Applied Optics},
  year = {2018},
  volume = {57},
  number = {18},
  pages = {4997},
  doi = {10.1364/AO.57.004997},
  url = {https://doi.org/10.1364/AO.57.004997}
}

@misc{Diffractsim2022,
  author = {de la Fuente Herrezuelo, Rafael},
  title = {{Diffractsim: A flexible Python diffraction simulator}},
  howpublished = {Zenodo},
  year = {2022},
  note = {\url{https://doi.org/10.5281/zenodo.6147771}},
  doi = {10.5281/zenodo.6147771},
  url = {https://doi.org/10.5281/zenodo.6147771}
}

@article{Mehta2014,
 author={Mehta, Shalin B. and Oldenbourg, Rudolf},
 title={{Image simulation for biological microscopy: microlith}},
 journal={Biomedical Optics Express},
 volume={5}, number={6}, pages={1822--1838}, year={2014},
 doi={10.1364/BOE.5.001822}
}

@article{CarterWolf1975,
 author={Carter, W. H. and Wolf, E.},
 title={{Coherence properties of lambertian and non-lambertian sources}},
 journal={Journal of the Optical Society of America},
 volume={65}, number={9}, pages={1067--1071}, year={1975}, doi={10.1364/JOSA.65.001067}
}

@article{Blomstedt2016,
 author = {Blomstedt, Kasimir and Set{\"a}l{\"a}, Tero and Tervo, Jani and Hoenders, Bernhard J. and Turunen, Jari and Friberg, Ari T.},
 title = {{Vector-valued Lambertian fields and their sources}},
 journal = {Physical Review A},
 year = {2016}, volume = {93}, number = {5}, pages = {053813},
 doi = {10.1103/PhysRevA.93.053813},
 url = {https://doi.org/10.1103/PhysRevA.93.053813}
}

@article{Peng2021,
 author = {Peng, Yifan and Choi, Suyeon and Kim, Jonghyun and Wetzstein, Gordon},
 title = {Speckle-free holography with partially coherent light sources and camera-in-the-loop calibration},
 journal = {Science Advances}, volume = {7}, number = {46}, pages = {eabg5040},
 year = {2021}, doi = {10.1126/sciadv.abg5040}
}

@article{Poulsen2025,
 author = {Poulsen, Kirstine Engell Sandager and Zambrana-Puyalto, Xavier and de la Fuente Herrezuelo, Rafael and Johansen, Villads Egede and Raza, S{\o}ren},
 title = {Experimental demonstration of a beam shaping non-imaging metasurface},
 journal = {Optics Express}, volume = {33}, number = {9}, pages = {19119--19129},
 year = {2025}, doi = {10.1364/OE.559542}
}

@inproceedings{KingmaBa2015,
  author = {Diederik P. Kingma and Jimmy Ba},
  title = {{Adam}: A method for stochastic optimization},
  booktitle = {International Conference on Learning Representations},
  year = {2015},
  url = {https://arxiv.org/abs/1412.6980}
}

@article{Schroder2011,
  author = {Schr{\"o}der, Sven and Duparr{\'e}, Angela and Coriand, Luisa and T{\"u}nnermann, Andreas and Penalver, Dayana H. and Harvey, James E.},
  title = {{Modeling of light scattering in different regimes of surface roughness}},
  journal = {Optics Express},
  year = {2011},
  volume = {19},
  number = {10},
  pages = {9820--9835},
  doi = {10.1364/OE.19.009820},
  url = {https://doi.org/10.1364/OE.19.009820}
}

@article{Gunoven2020,
  author = {G{\"u}n{\"o}ven, M. and Nasser, H. and {\"U}nal, M. and Aytekin, {\"O}. and Turan, R. and Bek, A.},
  title = {{Implementation of generalized Harvey-Shack theory in light scattering from rough surfaces}},
  journal = {Physical Review A},
  year = {2020},
  volume = {102},
  number = {6},
  pages = {063521},
  doi = {10.1103/PhysRevA.102.063521},
  url = {https://doi.org/10.1103/PhysRevA.102.063521}
}

@article{Ellingen2025,
  author = {Ellingen, Elias and Lange, Robert},
  title = {{Beam Shaping of VCSEL Arrays Using Randomized Microlens Arrays With Integrated Collimation}},
  journal = {IEEE Photonics Journal},
  year = {2025},
  volume = {17},
  number = {5},
  pages = {2201209},
  doi = {10.1109/JPHOT.2025.3603888},
  url = {https://doi.org/10.1109/JPHOT.2025.3603888}
}

@article{Kim2020,
  author = {Kim, Dong Cheon and Hermerschmidt, Andreas and Dyachenko, Pavel and Scharf, Toralf},
  title = {{Inverse design and demonstration of high-performance wide-angle diffractive optical elements}},
  journal = {Optics Express},
  year = {2020},
  volume = {28},
  number = {15},
  pages = {22321--22333},
  doi = {10.1364/OE.394442},
  url = {https://doi.org/10.1364/OE.394442}
}

@article{Wu2024,
  author = {Wu, Chenyang and Huang, Xuanlun and Ji, Yipeng and Cheng, Tingyu and Wang, Jiaxing and Chi, Nan and Yu, Shaohua and Chang-Hasnain, Connie J.},
  title = {{Addressable structured light system using metasurface optics and an individually addressable VCSEL array}},
  journal = {Photonics Research},
  year = {2024},
  volume = {12},
  number = {6},
  pages = {1129--1137},
  doi = {10.1364/PRJ.516942},
  url = {https://doi.org/10.1364/PRJ.516942}
}

@article{Khaidarov2020,
  author = {Khaidarov, Egor and Liu, Zhengtong and Paniagua-Dom{\'i}nguez, Ram{\'o}n and Ha, Son Tung and Valuckas, Vytautas and Liang, Xinan and Akimov, Yuriy and Bai, Ping and Png, Ching Eng and Demir, Hilmi Volkan and Kuznetsov, Arseniy I.},
  title = {Control of {LED} emission with functional dielectric metasurfaces},
  journal = {Laser \& Photonics Reviews},
  volume = {14},
  number = {1},
  pages = {1900235},
  year = {2020},
  doi = {10.1002/lpor.201900235}
}

@article{Sinha2026,
  author = {Sinha, Rajat Kumar and Mojahedi, Mo},
  title = {Nonlocal leaky mode metasurfaces for mode-, angle-, and polarization-multiplexed holography with thermal light},
  journal = {Laser \& Photonics Reviews},
  pages = {e72007},
  year = {2026},
  doi = {10.1002/lpor.72007}
}

\end{document}